\documentclass[12pt,letter]{article}
\usepackage[left=1in,right=1in,top=1.25in,bottom=1.25in]{geometry}

\usepackage{mathrsfs}
\usepackage{amssymb}
\usepackage{amsmath}
\usepackage{ascmac}
\usepackage{amsthm}
\usepackage[dvipdfmx]{graphicx}
\usepackage{booktabs}
\usepackage{enumerate}
\usepackage{setspace}
\usepackage{natbib}
\usepackage{color}
\usepackage{xcolor}
\usepackage{url}
\usepackage{comment}
\usepackage[colorlinks,linkcolor=blue,citecolor=blue,urlcolor=blue]{hyperref} 
\usepackage{times}

\usepackage{algpseudocode}
\usepackage{algorithm}

\newtheorem{thm}{Theorem}

\def\ep{{\varepsilon}}

\def\si{{\sigma}}

\def\beh{{\widehat \beta}}

\def\sih{{\widehat \si}}

\def\muh{{\widehat \mu}}

\def\bb{{\bar{b}}}
\def\bn{{\bar{n}}}

\def\bzeta{{\bar{\zeta}}}

\def\ba{{\text{\boldmath $a$}}}
\def\bA{{\text{\boldmath $A$}}}
\def\bbe{{\text{\boldmath $\beta$}}}
\def\bde{{\text{\boldmath $\delta$}}}
\def\bv{{\text{\boldmath $v$}}}
\def\bx{{\text{\boldmath $x$}}}
\def\by{{\text{\boldmath $y$}}}
\def\bz{{\text{\boldmath $z$}}}
\def\bX{{\text{\boldmath $X$}}}

\def\bZ{{\text{\boldmath $Z$}}}
\def\bb{{\text{\boldmath $b$}}}
\def\bR{{\text{\boldmath $R$}}}
\def\bI{{\text{\boldmath $I$}}}
\def\bSi{{\text{\boldmath $\Sigma$}}}

\def\bth{{\text{\boldmath $\theta$}}}

\def\bbt{{\widetilde{\bb}}}

\def\bbh{{\widehat{\bb}}}
\def\bthh{{\widehat{\bth}}}
\def\bbeh{{\widehat{\bbe}}}
\def\bRh{{\widehat{\bR}}}

\def\zero{{\text{\boldmath $0$}}}

\title{{\bf Robust Linear Mixed Models using Hierarchical Gamma-Divergence}}

\date{}

\begin{document}

\maketitle
\doublespacing

\vspace{-1.5cm}\noindent
\begin{center}
{\large
Shonosuke Sugasawa$^1$, Francis K. C. Hui$^2$ and A. H. Welsh$^2$\\
}
\vspace{0.5cm}
$^1$Faculty of Economics, Keio University\\
$^2$Research School of Finance, Actuarial Studies \& Statistics, The Australian National University

\vspace{0.7cm}
\noindent
{\large\bf Abstract}
\end{center}
Linear mixed models (LMMs) are a popular class of methods for analyzing longitudinal and clustered data. However, such models can be sensitive to outliers, and this can lead to biased inference on model parameters and inaccurate prediction of random effects if the data are contaminated. We propose a new approach to robust estimation and inference for LMMs using a hierarchical gamma-divergence, which offers an automated, data-driven approach to downweight the effects of outliers occurring in both the error and the random effects, using normalized powered density weights. For estimation and inference, we develop a computationally scalable minorization-maximization algorithm for the resulting objective function, along with a clustered bootstrap method for uncertainty quantification and a Hyvarinen score criterion for selecting a tuning parameter controlling the degree of robustness.
Under suitable regularity conditions, we show the resulting robust estimates can be asymptotically controlled even under a heavy level of (covariate-dependent) contamination. Simulation studies demonstrate hierarchical gamma-divergence consistently outperforms several currently available methods for robustifying LMMs. We also illustrate the proposed method using data from a multi-center AIDS cohort study.


\section{Introduction} \label{sec:intro}
Linear mixed models \citep[LMMs,][]{Ver2000} are one of the most widely used statistical methods for analyzing clustered or longitudinal data, with applications across many disciplines including medicine, agriculture and ecology. In the majority of these applications, practitioners will often make normality assumptions for both the random effects and the conditional distribution of the response (also known as the error terms). Along with an identity link, this results in a closed form marginal (multivariate normal distribution) log-likelihood function. As such, one can readily employ standard (restricted) maximum likelihood estimation and inference for model parameters, and construct predictions of the random effects. Many popular software packages for fitting LMMs e.g., the \texttt{lme4} \citep{Bates2015} and \texttt{glmmTMB} \citep{brooks2017glmmtmb} packages in \texttt{R}, are built around the assumption of joint normality for both the random effects and error terms.

It has long been recognized that in many real applications of LMMs, outliers can be present in both the random effects and error terms, with the amount of contamination sometimes quite severe. Under such circumstances, standard likelihood-based estimation such as is employed in the aforementioned software, can produce estimates very different from the true parameters and result in potentially poor statistical inference. 
This challenge has spurred a large body of statistical research into robustifying LMMs to handle situations where there is (strong \emph{a-priori} information concerning the presence of) outliers.
Generally speaking, robust LMMs fall into two classes of approaches: 1) modify the estimating equations or relevant likelihood functions by adjusting the residuals \citep[e.g.,][]{RW1995, Wang2005, Koller2016,zheng2021trimmed}; 2) replace the normality assumption for the random effects and/or error terms by other distributions such as the $t$-distribution or skew distributions \citep[e.g.,][]{Song2007,Lachos2010,maleki2019flexible,schumacher2021scale}.
Theoretical properties under the existence of a large number of extreme outliers are not necessarily well-established in the former approach, while in the latter approach the selection and estimation of the non-normal random effects distributions may be a challenging task methodologically and computationally. Moreover, as we will show in our simulation study in Section~\ref{sec:sim}, many existing approaches for robust LMMs do not provide reasonable estimates when the number of outliers or the distance between outliers and true model is relatively large.  

As an aside, we recognize that many methods have also been developed to address the related issue of diagnosing outliers and the understanding impacts of model misspecification in LMMs \citep[see][and references there in]{tanaka2020simple,Hui2020}, although this topic is outside the scope of this article.

There is a need to develop robust estimation and inferential procedures for LMMs that possess desirable theoretical properties, are computationally scalable with the number of clusters and cluster size, and produce strong finite sample performance in a variety of contamination settings. In this paper, we formulate a new method for robust estimation, inference, and prediction in independent cluster LMMs based on modifying the objective (likelihood) functions through the introduction of robust divergences. 
We begin by showing that the maximum likelihood estimator of the model parameters and predictors of the random effects, under the standard joint normality assumption of the response and error terms, can be characterized as the optimizer of a modified joint log-likelihood function which formulates the LMM in a hierarchical manner. 
We then propose to replace the components of this hierarchical formulation with $\gamma$-divergences \citep{Jones2001, FE2008}, leading to a new objective function for LMMs which we refer to as hierarchical $\gamma$-divergence. 

The $\gamma$-divergence has previously been developed as an alternative to the standard Kullback-Leibler divergence, and has has been considered in statistical models such as linear regression \citep{KF2017} and graphical modeling \citep{Hirose2017}. However, to our knowledge this paper is the first to introduce $\gamma$-divergence to mixed-effects modeling. Indeed, the proposed approach requires a novel use of dual $\gamma$-divergences to take into account outlier contamination at both levels of in (the hierarchical formulation of) the LMM. We also note recently, \cite{saraceno2023robust} applied a related density power divergence \citep{Basu1998} to the marginal distribution of LMMs. However they do not examine prediction of random effects, and in our numerical study we will show that our hierarchical $\gamma$-divergence almost always outperforms the density power divergence approach in terms of robustness across both estimation and prediction in LMMs.

We develop a computationally scalable algorithm based on the minorization-maximization (MM) algorithm \citep{HL2004} to fit LMMs using hierarchical $\gamma$-divergence. This iteratively updates the model parameters, random effects, and two types of density weights controlling the influence of outliers, and all of these steps possess fixed-point closed form expressions. As the $\gamma$-divergence involves a tuning parameter controlling the degree of robustness, we propose a data-driven method for choosing this based on the Hyvarinen score criterion of \citet{sugasawa2021selection}. Furthermore, for inference we employ a clustered bootstrap approach based on replacing the density weights in the MM-algorithm for fitting with modified weights that include a cluster-level random weight \citep[e.g.,][]{Field2010,o2018bootstrapping}. 
Under suitable regularity conditions as the number of clusters tends to infinity while the cluster sizes remain bounded, we show the proposed hierarchical $\gamma$-divergence estimators of the model parameters converge to fixed values, and that the distance between these fixed values and the true values \citep[known as the latent bias,][]{FE2008} depends on the degree of separability between the genuine and contamination distributions.
Hence, when the contamination distribution is sufficiently separated from the true model, the proposed estimator can be asymptotically controlled even under existence of an extreme level of outliers. 
Simulation studies demonstrate the proposed method typically outperforms many currently available robust LMM methods both in terms of estimation accuracy, statistical efficiency, and prediction, and under a wide range of outlier contamination scenarios. 

This rest of the article is organized as follows.
Section \ref{sec:HD} introduces independent cluster LMMs and an objective function that facilitates maximum likelihood estimation under joint normality assumptions, before detailing hierarchical $\gamma$-divergence for LMMs. 
Section \ref{sec:RLMM} develops the MM-algorithm to optimize the proposed divergence measure, the clustered bootstrap for statistical inference, and a selection method for the tuning parameter controlling robustness. In Section \ref{sec:robust}, we study the theoretical robustness properties of the proposed estimator, while Sections \ref{sec:sim} and \ref{sec:app} demonstrate the proposed method through numerical studies and an application to a longitudinal multi-center AIDS cohort study dataset, respectively. Finally, Section \ref{sec:dis} offers some avenues for future research. We provide \texttt{R} code for the proposed approach at \url{https://github.com/sshonosuke/Robust-LMM}, while all technical proofs are given in Supplementary Material.

\section{Linear Mixed Models and Hierarchical $\gamma$-Divergence}\label{sec:HD}

We focus on the independent cluster linear mixed model most commonly associated with clustered data analysis. Let $y_{ij}$ denote the $j$-th observation from the $i$-th cluster, where $j=1,\ldots,n_i$ and $i=1,\ldots,m$. We assume the cluster sizes are bounded above by some constant as $m \rightarrow \infty$. 
Next, let $\bx_{ij}$ denote a $p$-vector of covariates corresponding to the fixed effects, and $\bz_{ij}$ be a $q$-vector of covariates corresponding to the random effects. Both vectors may contain an intercept term as their first element, and we assume $p$ and $q$ are fixed.
The independent cluster LMM is formulated as 
\begin{equation}\label{LMM}
y_{ij}=\bx_{ij}^{\top}\bbe+\bz_{ij}^{\top}\bb_i+\ep_{ij},
\end{equation}
where $\bbe$ denotes the vector of fixed effect coefficients, $\bb_i$ denotes the vector of random effects of cluster $i$, and $\ep_{ij}$ denotes an error term. In many applications of LMMs, it is common to assume the random effects $\bb_i$ and error terms $\ep_{ij}$ are both normally distributed. That is, $\bb_i\sim N(\zero,\bR)$ and $\ep_{ij}\sim N(0,\si^2)$, where $\bR$ is an unstructured $q \times q$ random effects variance-covariance matrix and $\sigma^2$ is the error variance. Conditional on $\bb_i$ then, the responses $y_{ij}$ are assumed to be independent observations from a normal distribution. 
Note all the developments below can be extended to the case where the independence assumption for the error terms is relaxed e.g., the $\ep_{ij}$'s within a cluster are correlated. However, given the commonality of assuming independent errors, conditional on the random effects, in many applications of linear mixed models \citep[see for instance][among others]{verbeke1997effect,zhang2001linear,lin2013fixed,saraceno2023robust}, then we will focus on case of independent error terms below. We denote the vector of model parameters in equation~(\ref{LMM}) as $\bth=(\bbe^{\top}, \sigma^2, {\rm vech}(\bR)^{\top})^{\top}$ where ${\rm vech}(\cdot)$ denotes the half-vectorization operator. 
Also, we let $\bb=(\bb_1^{\top},\ldots,\bb_m^{\top})^{\top}$ denote the full $mq$-vector of random effects.

In practice, the most common method for estimating LMMs, and which forms the basis of our developments in this article, is maximum likelihood estimation (or its restricted variant) for which the marginal log-likelihood function 
\begin{equation}\label{mL}
L_M(\bth) = \sum_{i=1}^m\log\phi_{n_i}(\by_i; \bX_i\bbe, \bSi_i),
\end{equation}
where $\by_i=(y_{i1},\ldots,y_{in_i})^{\top}$, $\bX_i=(x_{i1},\ldots,x_{in_i})^{\top}$, and $\bSi_i=\bZ_i\bR\bZ^\top_i+\sigma^2\bI_{n_i}$ denotes the marginal covariance matrix of $\by_i$ with $\bZ_i=(z_{i1},\ldots,z_{in_i})^{\top}$ and $\bI_{n_1}$ representing an identity matrix of dimension $n_i$. 
Also, we let $\phi_{n_i}(\cdot; \ba, \bA)$ denote the $n_i$-variate normal density with mean vector $\ba$ and covariance matrix $\bA$. 
After maximum likelihood estimation of $\bth$, the random effects are typically predicted through the conditional expectation of $\bb_i$ given $\by_i$, also known as the Best Linear Unbiased Predictor or BLUP \citep[e.g.,][]{robinson1991blup,lyu2022asymptotics}, which takes the form $\bbt_i(\bth)=\bR\bZ_i^\top \bSi_i^{-1}(\by_i-\bX_i\bbe)$. 
Let $\bthh_{\rm ML}={\rm argmax}_{\bth}L_M(\bth)$ and $\bbh_{\rm ML}=\bbt(\bthh_{\rm ML})$ denote the resulting maximum likelihood estimator of $\bth$, and empirical BLUP for the full vector of random effects $\bb$ in the LMM, respectively.

As a starting point for our developments in robustifying LMMs, we note that both the maximum likelihood estimators and empirical BLUPs can be obtained by considering the following single, modified joint log-likelihood function, 
\begin{align}\label{JL}
L_J(\bth, \bb) = &\sum_{i=1}^m\sum_{j=1}^{n_i}\log \phi(y_{ij};\bx_{ij}^{\top}\bbe+\bz_{ij}^{\top}\bb_i,\sigma^2)+\sum_{i=1}^m\log \phi_q(\bb_i;\zero,\bR)\nonumber \\
&-\frac12\sum_{i=1}^m\log\det(\bSi_i)
+\frac{N}{2}\log(\sigma^2)+\frac{m}{2}\log\det(\bR).
\end{align}
The first two terms in $L_J(\bth, \bb)$ correspond to the joint log-likelihood function of the response and random effects, while the remaining three terms ensure that profiling $\bb$ out from $L_J(\bth, \bb)$ leads to the marginal likelihood $L_M(\bth)$. 
We can then show that $\bthh_{\rm ML}$ and $\bbh_{\rm ML}$ jointly maximize equation (\ref{JL}), and refer the reader to Supplementary Material for a proof of this.

In the modified joint log-likelihood function, the hierarchy of the LMM is evident through the explicit form for the conditional distribution of the observations given the random effects, and the marginal distribution of the random effects. In the next section, we shall see that this hierarchical formulation for obtaining $\bthh_{\rm ML}$ and $\bbh_{\rm ML}$ is central to motivating a hierarchical divergence that achieves robustness for estimating the model parameters and the random effects of the LMM.

\subsection{Hierarchical $\gamma$-divergence}
We propose replacing each of the two log-likelihood components in $L_J(\bth, \bb)$, characterizing the hierarchical formulation of the LMM, by robust objective functions based on $\gamma$-divergence \citep{Jones2001,FE2008}. Briefly, for an independent but not necessarily identically distributed sample, $y_1,\ldots,y_n$, the $\gamma$-divergence for a statistical model $f(\cdot;\bth)$ admits the general definition 
$$
\frac{n}\gamma\log\left(\frac1n\sum_{i=1}^nf(\by_i;\bth)^{\gamma}\right) -\frac{n}{1+\gamma}\log\left(\frac1n\sum_{i=1}^n\int f(t;\bth)^{1+\gamma}dt\right),
$$
which, as $\gamma\to 0$, the above converges to the standard log-likelihood $\sum_{i=1}^n \log f(\by_i;\bth)$. The tuning parameter $\gamma > 0$ controlling the degree of robustness, such that large values of $\gamma$ will downweight outlying observations more severely; see the next section for more discussion of this. 
Returning to the LMM, since the $y_{ij}$'s are conditionally independent given $\bb_i$, and the $\bb_i$'s are also marginally independent of each other, we can apply the above divergence measure to both the distribution of the response $\phi(y_{ij};\bx_{ij}^{\top}\bbe+\bz_{ij}^{\top}\bb_i,\sigma^2)$ and the distribution of the random effects $\phi_q(\bb_i;\zero,\bR)$ separately. 
Following straightforward algebra, this results in a modification of equation \eqref{JL} for robustifying LMMs, which we refer to as the hierarchical $\gamma$-divergence:
\begin{equation}\label{Q}
\begin{split}
D_{\gamma}(\bth,\bb) 
&=\frac{N}\gamma\log\left(\frac1N\sum_{i=1}^m\sum_{j=1}^{n_i}\phi(y_{ij};\bx_{ij}^{\top}\bbe+\bz_{ij}^{\top}\bb_i,\sigma^2)^{\gamma}\right) +\frac{N(1+2\gamma)}{2(1+\gamma)}\log(\sigma^2) \\
&
+\frac{m}\gamma\log\left(\frac1m\sum_{i=1}^m\phi_q(\bb_i;\zero,\bR)^{\gamma}\right)+\frac{m(1+2\gamma)}{2(1+\gamma)}\log\det(\bR)
-\frac12\sum_{i=1}^m\log\det(\bSi_i). 
\end{split}
\end{equation}
We denote the new robust estimators as $(\bthh_\gamma,\bbh_\gamma) = \arg\max_{\bb,\bth} D_{\gamma}(\bth,\bb)$. As hinted at above, we can show that $
\lim_{\gamma\to 0}\left(D_{\gamma}(\bth,\bb) - (N+m)/\gamma\right) = L_J(\bth,\bb)$, i.e., the robust estimators $\bbh_\gamma$ and $\bthh_\gamma$ will almost be equivalent to the standard maximum likelihood estimator $\bbh_{\rm ML}$ and empirical BLUP $\bthh_{\rm ML}$ for small values of $\gamma$. The tuning parameter in \eqref{Q} thus controls the robustness of the hierarchical $\gamma$-divergence against outliers in both levels of the LMM. 
Larger values of $\gamma$ corresponding to downweighting the relative contributions of outlying error terms and random effects more strongly. 
We will discuss a data-dependent selection of $\gamma$ in Section \ref{sec:selection}.

\subsection{Connections to weighted estimating equations} \label{subsec:WEEconnection}
The proposed hierarchical $\gamma$-divergence can be examined from the viewpoint of estimating equations from LMMs. 
For instance, we can derive the score equations of  (\ref{Q}) with respect to $\bbe$ and $\bb_i$ as
\begin{equation}\label{WEE}
\begin{split}
&\frac1{N}\sum_{i=1}^m\sum_{j=1}^{n_i}w_{ij}\bx_{ij}(y_{ij}-\bx_{ij}^{\top}\bbe-\bz_{ij}^{\top}\bb_i)=\zero,\\
&\frac1{\sigma^2}\sum_{j=1}^{n_i}w_{ij}\bz_{ij}(y_{ij}-\bx_{ij}^{\top}\bbe-\bz_{ij}^{\top}\bb_i)-u_i\bR^{-1}\bb_i=\zero, 
\end{split}
\end{equation}
where $w_{ij}= w(y_{ij}; \bb_i,\bbe,\sigma^2)$ and $u_i= u(\bb_i; \bR)$ denote normalized model-based density weights defined as
\begin{equation}\label{weight}
\begin{split}
w(y_{ij}; \bb_i,\bbe,\sigma^2) &= \frac{\phi(y_{ij};\bx_{ij}^{\top}\bbe+\bz_{ij}^{\top}\bb_i,\sigma^2)^{\gamma}}{N^{-1}\sum_{i=1}^m\sum_{j=1}^{n_i}\phi(y_{ij};\bx_{ij}^{\top}\bbe+\bz_{ij}^{\top}\bb_i,\sigma^2)^{\gamma}}, \\
u(\bb_i; \bR) &= \frac{\phi(\bb_i;\zero,\bR)^{\gamma}}{m^{-1}\sum_{i=1}^m\phi(\bb_i;\zero,\bR)^{\gamma}},
\end{split}
\end{equation}
such that $\sum_{i=1}^m\sum_{j=1}^{n_i}w_{ij}=N$ and $\sum_{i=1}^mu_i=m$. 
It is clear from the above forms that, assuming the mean structure for the LMM is correctly specified, the weights for relatively outlying clusters/observations are down-weighted for $\gamma>0$, such that their effect in the estimation process is limited in hierarchical $\gamma$-divergence. For instance, if $\ep_{ij}$ is an outlier then the conditional density of the response $\phi(y_{ij};\bx_{ij}^{\top}\bbe+\bz_{ij}^{\top}\bb_i,\sigma^2)$ will be relatively small (assuming the parameters are close to the true values), which implies $w_{ij}$ will also be small relative to less outlying error terms.
Analogously, to understand the role of the weights $u_i$, we consider the score equation for $\bR$ derived from (\ref{Q}) as 
$$
-\frac12\sum_{i=1}^m u_{i}\bR^{-1}\bb_i\bb_i^{\top}\bR^{-1}+\frac{m\gamma}{2(1+\gamma)}\bR^{-1}-\frac12\sum_{i=1}^m\bZ_i^\top\bSi_i^{-1}\bZ_i=\zero.
$$
When the random effect $\bb_i$ is outlying, then the corresponding weight $u_i$ will be small and such outlying random effects will be downweighted in the estimation of the random effects variance-covariance matrix.
This capacity to separately but jointly robustify the two levels of the LMM is a key aspect our proposed hierarchical $\gamma$-divergence.
Also, while for simplicity we have used the same value of $\gamma$ at both levels of the LMM, this can be generalized to use two separate values of $\gamma$ in the hierarchical formulation. 
For example, if one wishes to apply the $\gamma$-divergence only to the errors while leaving the random effects unchanged, such a setting can be viewed as a special case of the hierarchical $\gamma$-divergence method.

It is important to acknowledge that the use of weighted estimating equations for robust estimation has been proposed before for LMMs. As reviewed in Section \ref{sec:intro}, \cite{Koller2016} for example employed Huberized versions of the ``residual" terms in (\ref{WEE}), although the resulting method was computationally quite burdensome and no theoretical properties were established. By contrast, the hierarchical $\gamma$-divergence estimator can be efficiently computed via a iterative algorithm which exploits the weighted estimating equations form above, while  its robustness properties are theoretically established in Section~\ref{sec:robust}.

\section{Robust Estimation and Inference}
\label{sec:RLMM}

\subsection{Estimation algorithm}
In this section, we develop a computationally scalable method for maximizing the hierarchical $\gamma$-divergence via the Minorization-Maximization (MM) algorithm \citep{HL2004}. 
Let $\bth^{(t)}$ and $\bb^{(t)}$ denote the values of $\bth$ and $\bb$, respectively, at the $k$-th iteration of the MM-algorithm, and define the corresponding weights as $w^{(t)}_{ij}=w(y_{ij}; \bb^{(t)}_{i},\bbe^{(t)},\sigma^{(t)2})$ and $u^{(t)}_{i}=u(\bb^{(t)}_{i}; \bR^{(t)})$, using the forms given in (\ref{weight}). We can then apply Jensen's inequality $\log\left(\sum_{i=1}^N a_ic_i \right)\geq\sum_{i=1}^N a_i\log(c_i)$ for $a_i,c_i>0$ satisfying $\sum_{i=1}^N a_i = 1$, to minorize the first term in $D_{\gamma}(\bth,\bb)$.
Specifically, by setting $a_{ij}=w^{(t)}_{ij}/N$ and $c_{ij} = N\phi(y_{ij};\bx_{ij}^\top \bbe + \bz_{ij}^\top \bb_i, \sigma^2)^{\gamma}/w^{(t)}_{ij}$ \citep[see also][]{KF2017}, it holds that 
$$
\log\left(\sum_{i=1}^m \sum_{j=1}^{n_i}\phi(y_{ij};\bx_{ij}^\top \bbe + \bz_{ij}^\top \bb_i)^{\gamma}\right)
=\log\left(\sum_{i=1}^m \sum_{j=1}^{n_i} a_{ij}c_{ij}\right) \geq 
\sum_{i=1}^m \sum_{j=1}^{n_i} a_{ij}\log(c_{ij}),
$$
which gives the lower bound of the first term in $D_{\gamma}(\bth,\bb)$.
An analogous approach can be applied to minorize the third term in $D_{\gamma}(\bth,\bb)$. 
On combining these two results, we obtain the following minorization function of $D_{\gamma}(\bth,\bb)$ at the $t$-th iteration:
\begin{align*}
D_{\gamma}(\bth,\bb)&\geq \frac{1}\gamma\sum_{i=1}^m\sum_{j=1}^{n_i}w^{(t)}_{ij}\log\left(\frac{N\phi(y_{ij};\bx_{ij}^\top \bbe + \bz_{ij}^\top \bb_i, \sigma^2)^{\gamma}}{w^{(t)}_{ij}}\right)
+\frac{N(1+2\gamma)}{2(1+\gamma)}\log(\sigma^2)\notag \\
&\quad +\frac{1}\gamma\sum_{i=1}^m u^{(t)}_{i}\log\left(\frac{m\phi_q(\bb_i; \zero,\bR)^{\gamma}}{u^{(t)}_{i}}\right)
+\frac{m(1+2\gamma)}{2(1+\gamma)}\log\det(\bR)-\frac12\sum_{i=1}^m\log\det(\bSi_i) \notag\\
&=-\frac{1}{2\sigma^2}\sum_{i=1}^m\sum_{j=1}^{n_i}w^{(t)}_{ij}(y_{ij}-\bx_{ij}^{\top}\bbe-\bz_{ij}^{\top}\bb_i)^2-\frac12\sum_{i=1}^mu^{(t)}_{i}\bb_i^{\top}\bR^{-1}\bb_i\\
&\quad + \frac{N\gamma}{2(1+\gamma)}\log(\sigma^2)+\frac{m\gamma}{2(1+\gamma)}\log\det(\bR)-\frac12\sum_{i=1}^m\log\det(\bSi_i)+C \\
&\equiv D_{\gamma, M}^{(t)}(\bth, \bb),
\end{align*}
where $C$ is a constant as a function of $\bth$ and $\bb$, and equality is obtained if and only if $\bth = \bth^{(t)}$ and $\bb = \bb^{(t)}$. 
It is straightforward to see $D_{\gamma, M}^{(t)}(\bth, \bb)$ is a quadratic function of $\bbe$ and $\bb_i$, and indeed it can be viewed as a weighted version of the hierarchical $\gamma$-divergence in (\ref{JL}). More importantly, it follows that we can maximize $D_{\gamma, M}^{(t)}(\bth, \bb)$, or equivalently solve a set of weighted estimating equations similar to equations \eqref{WEE}-\eqref{weight}, to obtain a set of closed-form updates for $\bbe$ and $\bb_i$; see Algorithm \ref{alg:MMalgorithm} for details. For the random effects covariance matrix and the error variance, the corresponding first order derivatives of the minorization function are given by 
\begin{align*}
&\frac{\partial D_{\gamma, M}^{(t)}(\bth, \bb)}{\partial\sigma^2}=\frac{1}{2\sigma^4}\sum_{i=1}^m\sum_{j=1}^{n_i}w^{(t)}_{ij}(y_{ij}-\bx_{ij}^{\top}\bbe-\bz_{ij}^{\top}\bb_i)^2+\frac{N\gamma}{2(1+\gamma)\sigma^2}-\frac12\sum_{i=1}^m{\rm tr}(\bSi_i^{-1})\\
&\frac{\partial D_{\gamma, M}^{(t)}(\bth, \bb)}{\partial\bR}=-\frac12\sum_{i=1}^mu^{(t)}_{i}\bR^{-1}\bb_i\bb_i^{\top}\bR^{-1}+\frac{m\gamma}{2(1+\gamma)}\bR^{-1}-\frac12\sum_{i=1}^m\bZ_i^\top\bSi_i^{-1}\bZ_i,
\end{align*}
from which a closed-form fixed-point iterative update can be computed for $\sigma^2$ and $\bR$. A formal algorithm for fitting the LMM using the hierarchical $\gamma$-divergence is provided in Algorithm~\ref{alg:MMalgorithm}.

\begin{algorithm}[tb] 
\caption{LMM estimation using hierarchical $\gamma$-divergence}
\label{alg:MMalgorithm}
\begin{algorithmic}

\Require Initial estimates and predictions $\bth^{(0)}$, $\bb^{(0)}$; tolerance value $e_{\rm tol}$ e.g., $e_{\rm tol}=10^{-6}$.

\State $t \gets 0$ and $d \gets 1$
\While{$d> e_{\rm tol}$}
\State 
\underline{Update weights (see equation (\ref{weight}) for specific form):}
$$
w_{ij}^{(t)} \gets 
w(y_{ij}; \bb_{i}^{(t)},\bbe^{(t)},\sigma^{(t)2}); \quad u_i^{(t)}=u(b_{i}^{(t)}; \bR^{(t)}).
$$

\State 
\underline{Update parameters and random effects:}
\begin{equation}\label{update}
\begin{split}
&\bbe^{(t+1)}
\gets\bigg(\sum_{i=1}^m\sum_{j=1}^{n_i}w_{ij}^{(t)}\bx_{ij}\bx_{ij}^{\top}\bigg)^{-1}\sum_{i=1}^m\sum_{j=1}^{n_i}w_{ij}^{(t)}\bx_{ij}\Big(y_{ij}-\bz_{ij}^{\top}\bb_i^{(t)}\Big),\\
&b_{i}^{(t+1)}
\gets\bigg(\sum_{j=1}^{n_i}w_{ij}^{(t)}\bz_{ij}\bz_{ij}^{\top}+u_i^{(t)}\sigma^{(t)2}(\bR^{(t)})^{-1}\bigg)^{-1}\sum_{j=1}^{n_i}w_{ij}^{(t)}\bz_{ij}\Big(y_{ij}-\bx_{ij}^{\top}\bbe^{(t+1)}\Big),\\
&\sigma^{(t+1)2}
\gets\frac
{\sum_{i=1}^m\sum_{j=1}^{n_i}w_{ij}^{(t)}\Big(y_{ij}-\bx_{ij}^{\top}\bbe^{(t+1)}-\bz_{ij}^{\top}\bb_i^{(t+1)}\Big)^2}
{\sigma^{(t)2}\sum_{i=1}^m{\rm tr}(\bSi_i^{(t)-1})-N\gamma/(1+\gamma)},\\
&\bR^{(t+1)}
\gets \frac{1+\gamma}{m} \left(\sum_{i=1}^mu_i^{(t)}\bb_i^{(t+1)}\bb_i^{(t+1)\top} - \bR^{(t)}\sum_{i=1}^m\bZ_i^\top\bSi_i^{(t)-1}\bZ_i\bR^{(t)} + m\bR^{(t)}\right).
\end{split}
\end{equation} 

\State
$d \gets |D(\bth^{(t+1)}, \bb^{(t+1)})-D(\bth^{(t)}, \bb^{(t)})|; \quad t\gets t+1$
\EndWhile

\Ensure $\bthh=\bth^{(t+1)}$ and $\bbh=\bb^{(t+1)}$ 
\end{algorithmic}
\end{algorithm}

It can be seen that the proposed estimation approach for hierarchical $\gamma$-divergence is computationally more scalable than some of the existing robust methods for LMMs \citep[e.g.,][ requires some form of Monte Carlo integration with respect to the random effects]{Lachos2010,Koller2016}. 
Also, note that as an MM-algorithm, we are guaranteed an ascent property, i.e., the value of the $D_{\gamma}(\bth,\bb)$ will monotonically increase with each iteration. This is not necessarily the case with other robust approaches to fitting LMMs.

\subsection{A clustered bootstrap for inference}\label{sec:inference}
For inference and uncertainty quantification in the LMM, we propose to use the clustered bootstrap of \citep{Field2010,o2018bootstrapping}, which is a special case of the generalized bootstrap for estimating equations. 
We highlight a bootstrap approach is computationally reasonable for LMMs here, owing to the simplicity of the estimation algorithm for hierarchical $\gamma$-divergence; see also \cite{Chat2005} and \cite{Field2010} and example theoretical justifications of the clustered bootstrap, although we leave a more formal proof of  consistency in the context of the hierarchical $\gamma$-divergence approach specifically as an avenue of future research.

In the clustered bootstrap, we incorporate a set of cluster-specific random weights $\xi=(\xi_1,\ldots,\xi_m)^\top$ in the relevant estimating equations for the LMM, noting the hierarchical $\gamma$-divergence already bears close connections to weighted estimating equations for robustness as discussed in Section \ref{subsec:WEEconnection}. In detail, first define $w^{\dagger}_{ij}=\phi(y_{ij}; \bx_{ij}^{\top}\bbe+\bz_{ij}^{\top}\bb_i, \sigma^2)^{\gamma}$, which is the unnormalized version of the first model-based density weight in equation \eqref{weight}. Then for the fixed effect coefficients $\bbe$, we can incorporate the random weight $\xi_i$ to obtain
\begin{align*}
\frac{\sum_{i=1}^m\xi_i\sum_{j=1}^{n_i}w^{\dagger}_{ij}(y_{ij}-\bx_{ij}^{\top}\bbe-\bz_{ij}^{\top}\bb_i)}{\sum_{i=1}^m\xi_i\sum_{j=1}^{n_i}w^{\dagger}_{ij}}=0,
\end{align*}
from which we can see that this is equivalent to replacing the normalized weights $w_{ij}$ with a modified but still normalized bootstrap weight $w_{ij}^{\xi}= \xi_iw^{\dagger}_{ij}/\sum_{i=1}^m\xi_i\sum_{j=1}^{n_i}w^{\dagger}_{ij}$ in equation (\ref{WEE}).
The same argument can be applied to the estimating equation for $\sigma^2$, for which the form of the score $\partial D_{\gamma, M}(\bth, \bb)/\sigma^2$ was presented earlier, while in the estimating equation for $\bR$ it can be seen that introducing random weights $\xi_i$ is the same as replacing $u_i$ by $u_i^{\xi} = \xi_iu^{\dagger}_i/\sum_{i=1}^m\xi_i u^{\dagger}_i$ where $u^{\dagger}_i=\phi(\bb_i;\zero,\bR)^{\gamma}$.
Finally, turning to the random effects, we first rewrite the second equation of (\ref{WEE}) in the form of the summation over all the clusters.
By multiplying $\sum_{i=1}^m u_i^{\dagger}$ and $\sum_{i=1}^m\sum_{j=1}^{n_i}w_{ij}^{\dagger}$ by the second equation of (\ref{WEE}), we have 
$$
\sum_{i'=1}^m\left(\frac{u_{i'}^{\dagger}}{\sigma^2}\sum_{j=1}^{n_i}w_{ij}^{\dagger}(y_{ij}-\bx_{ij}^{\top}\bbe-\bz_{i j}^{\top}\bb_i)-u_i^{\dagger}\bR^{-1}b_i\sum_{j=1}^{n_{i'}}w_{i'j}^{\dagger}\right) = 0, \ \ \ i=1,\ldots,m.
$$
Then, a randomized-weighted version can be obtained by replacing the sum over $i$ with the weighted sum using the quantities defined above. 
This results in  
$$
\sum_{i'=1}^m\xi_{i'}\left(\frac{u_{i'}^{\dagger}}{\sigma^2}\sum_{j=1}^{n_i}w_{ij}^{\dagger}(y_{ij}-\bx_{ij}^{\top}\bbe-\bz_{i j}^{\top}\bb_i)-u_i^{\dagger}\bR^{-1}b_i\sum_{j=1}^{n_{i'}}w_{i'j}^{\dagger}\right) = 0, \ \ \ i=1,\ldots,m,
$$
where $\xi_{i'}$ is a random weight. 
By multiplying through by $\xi_i$, the above equation can be rewritten as 
$$
\sum_{i'=1}^m\left(\frac{\xi_{i'} u_{i'}^{\dagger}}{\sigma^2}\sum_{j=1}^{n_i}\xi_iw_{ij}^{\dagger}(y_{ij}-\bx_{ij}^{\top}\bbe-\bz_{i j}^{\top}\bb_i)-\xi_iu_i^{\dagger}\bR^{-1}b_i\sum_{j=1}^{n_{i'}}\xi_{i'}w_{i'j}^{\dagger}\right) = 0, \ \ \ i=1,\ldots,m,
$$
which is a function of $\xi_iu_i^{\dagger}$ and $\xi_iw_{ij}^{\dagger}$. It follows that the above is equivalent to replacing $w_{ij}$ and $u_i$ by $w_{ij}^{\xi}$ and $u_i^{\xi}$, respectively, in the weighted estimating equation for $\bb_i$ in (\ref{WEE}).

To summarize, applying the clustered bootstrap for hierarchical $\gamma$-divergence straightforwardly involves replacing the weights $w_{ij}$ and $u_i$ in the weighted estimating equations by modified normalized bootstrapped weights $w_{ij}^{\xi}$ and $u_i^{\xi}$. Moreover, it is not hard to show that all of the above is equivalent to maximizing the weighted version of the hierarchical $\gamma$-divergence measure
\begin{equation*}
\begin{split}
D_{\gamma}^{\xi}(\bth,\bb) 
&\equiv \frac{N}\gamma\log\left(\frac1N\sum_{i=1}^m\xi_i\sum_{j=1}^{n_i}\phi(y_{ij};\bx_{ij}^{\top}\bbe+\bz_{ij}^{\top}\bb_i,\sigma^2)^{\gamma}\right) +\frac{N(1+2\gamma)}{2(1+\gamma)}\log(\sigma^2) \\
&\ \ \ \ 
+\frac{m}\gamma\log\left(\frac1m\sum_{i=1}^m\xi_i\phi_q(\bb_i;\zero,\bR)^{\gamma}\right) + \frac{m(1+2\gamma)}{2(1+\gamma)}\log\det(\bR)
-\frac12\sum_{i=1}^m\log\det(\bSi_i), 
\end{split}
\end{equation*}
which can be straightforwardly maximized again using an MM algorithm analogous to Algorithm \ref{alg:MMalgorithm}.
Finally, regarding the choice of random weights, we set $\xi \sim m\times {\rm Dir}(1,\ldots,1)$ such that the expectation of $E(\xi_i) = 1$, although other distribution are also possible. By randomly sampling the random weights a large number of times e.g. we used 500 bootstrap datasets in all the simulations and application, and applying the MM-algorithm to each, we  obtain a set of bootstrap estimates of the model parameters and predictions of random effects. From this, bootstrap confidence/predictions intervals for all model parameters and random effects, respectively, can be constructed via the percentile method.

\subsection{Tuning parameter selection}\label{sec:selection}
The tuning parameter $\gamma > 0$ controls the degree of robustness in the hierarchical $\gamma$-divergence; if $\gamma$ is large, the normalized density weights for outliers  decrease rapidly so outliers are down-weighted more severely. However as with all robust estimation procedures, there is a natural trade-off as using larger values of $\gamma$ leads to statistically inefficient estimation and prediction when few or no outliers exist in the data \citep{Basu1998,Jones2001}. As a data-driven approach to tuning $\gamma$ then, we adopt the criterion of \cite{sugasawa2021selection} based on the Hyvarinen score \citep[H-score,][]{hyvarinen2005estimation}.

Briefly, given a set of independent observations $y_1,\ldots,y_n$ and a statistical model $f(y;\bth)$ for the observations, the H-score is defined as
$
\sum_{i=1}^n(2 \partial^2 \log f(y_i;\bth) / \partial^2 y_i + (\partial\log f(y_i;\bth) / \partial y_i)^2),
$
where we note that the derivatives are with respect to the observations and not the model parameters.
We also note that the H-score for multivariate data can be defined in a similar way.
Minimizing the H-score is equivalent to minimizing the Fisher divergence of a model $f(y;\bth)$ and the data generating distribution. 
\cite{sugasawa2021selection} applied the H-score to a pseudo-statistical model obtained by a transformation of the $\gamma$-divergence measure, and we employ a similar approach here for the independent cluster LMM. Specifically, since equation \eqref{Q} involves two $\gamma$-divergences for the conditional distribution of the response and the marginal distribution of the random effects, we compute two separate H-scores as follows: 
\begin{itemize}
\item The H-score associated with the response/error terms of the LMM is given by 
\begin{equation*}\label{H1}
H_1(\gamma) = \sum_{i=1}^m\sum_{j=1}^{n_i}\left(\frac{2\left(\gamma(y_{ij}-\muh_{ij\gamma})^2-\sih_{\gamma}^2\right)}{\sih_{\gamma}^4 C_{\gamma}(\sih_{\gamma}^2)}\phi(y_{ij};\muh_{ij\gamma},\sih_{\gamma}^2)^{\gamma}
+
\frac{(y_{ij}-\muh_{ij\gamma})^2}{\sih_{\gamma}^4C_{\gamma}(\sih_{\gamma}^2)^2}\phi(y_{ij};\muh_{ij\gamma},\sih_{\gamma}^2)^{2\gamma}\right),    
\end{equation*}
where $\muh_{ij\gamma} = \bx_{ij}^{\top}\bbeh_{\gamma}+\bz_{ij}^{\top}\bbh_{i\gamma}$ and $C_{\gamma}(\sigma^2) = \{(1+\gamma)^{-1/2}(2\pi\sigma^2)^{-\gamma/2}\}^{\gamma/(1+\gamma)}$.

\item The H-score associated with the random effects is given by 
\begin{equation*}\label{H2}
H_2(\gamma)=\sum_{i=1}^m\left(\frac{2\{\gamma(1_q\bRh_\gamma^{-1}\bbh_{i\gamma})^2-{\rm tr}(\bRh_\gamma^{-1})\}}{C_{\gamma}(\bRh_{\gamma})}\phi(\bbh_{i\gamma};\zero,\bRh_{\gamma})^{\gamma}
+
\frac{(1_q\bRh_\gamma^{-1}\bbh_{i\gamma})^2}{C_{\gamma}(\bRh_{\gamma})^2}\phi(\bbh_{i\gamma};\zero,\bRh_{\gamma})^{2\gamma}\right),
\end{equation*}
where $C_{\gamma}(\bR) = \{(1+\gamma)^{-q/2}(2\pi)^{-q\gamma/2}\det(\bR)^{-\gamma/2}\}^{\gamma/(1+\gamma)}$. 
\end{itemize}

We propose to choose $\gamma$ by first computing $\gamma_{\rm opt}^{(1)}={\rm argmin}_{\gamma\in\Gamma}H_1(\gamma)$ and $\gamma_{\rm opt}^{(2)}={\rm argmin}_{\gamma\in\Gamma}H_2(\gamma)$, where $\Gamma=\{\gamma_1,\ldots,\gamma_L\}$ is a candidate set. 
Afterward, we set $\gamma_{\rm opt}=\max(\gamma_{\rm opt}^{(1)},\gamma_{\rm opt}^{(2)})$. Note the tuning parameter is selected only once, and we use the same value throughout the clustered bootstrap procedure discussed in Section \ref{sec:inference}.
Further, the above procedure can be extended to use different $\gamma$ for the response and random effects, although the computation cost will increase significantly.

\section{Robustness property}\label{sec:robust}
We examine the theoretical properties of the hierarchical $\gamma$-divergence estimator for LMMs under outlier contamination, focusing on the model parameters. 
To this end, we first note the robust estimator $\bthh_{\gamma}$ from (\ref{Q}) can be equivalently obtained from a profile-like hierarchical $\gamma$-divergence, $\bthh_{\gamma} = {\rm argmax}_{\bth}D_{\gamma}^M(\bth) = 
{\rm argmax}_{\bth} D_{\gamma}(\bth, \bbt(\bth))$,
where $\bbt(\bth) = (\bbt_1(\bth)^\top,\ldots,\bbt_m(\bth)^\top)$ is the solution of the second weighted estimating equation given in (\ref{WEE}). 
Let $\bth^{\ast}=(\bbe^{\ast\top}, \sigma^{\ast 2}, {\rm vech}(\bR^{\ast})^{\top})^{\top}$ denote the true parameter point of $\bth$, and define the uncontaminated marginal distribution of $\by_i$ for the LMM as $\phi_{n_i}(\by_i; \bX_i^\top\bbe^{\ast}, \bZ_i\bR^{\ast}\bZ_i^\top + \sigma^{\ast 2}\bI_{n_i}) \equiv G(\by_i;\bth^{\ast}); i =1,\ldots,m$. 
Given $\gamma$, we now define the pseudo-true parameter $\bth_{\gamma}^{\ast}$ as the maximizer of $\lim_{m\to \infty}m^{-1}D_{\gamma}^{M\ast}(\bth)$, where 
\begin{equation*}
\begin{split}
\frac{1}{m} D_{\gamma}^{M\ast}(\bth) 
&= \frac{\bn}{\gamma}\log\left(\frac1{m\bn}\sum_{i=1}^m\sum_{j=1}^{n_i}\int\phi(y_{ij};\bx_{ij}^{\top}\bbe+\bz_{ij}^{\top}\bbt_i(\by_i;\bth),\sigma^2)^{\gamma}G(\by_i;\bth^{\ast})dy_{ij}\right) \\
&
+\frac{1}\gamma\log\left(\frac1m\sum_{i=1}^m\int\phi_q(\bbt_i(\by_i;\bth);\zero,\bR)^{\gamma}G(\by_i;\bth^{\ast})dy_{ij}\right) +C_{\gamma}(\bth), 
\end{split}
\end{equation*}
with $\bar{n}=\lim_{m\to\infty}m^{-1}\sum_{i=1}^m n_i$ as the limiting average sample size, and 
$
C_{\gamma}(\bth) = (1+2\gamma)/(2(1+\gamma))(\bn\log(\sigma^2) + \log\det(\bR)) - (2m)^{-1}\sum_{i=1}^m\log\det(\bSi_i).
$
The quantity $\lim_{m\to \infty}m^{-1}D_{\gamma}^{M\ast}(\bth)$ can be interpreted as the limiting hierarchical $\gamma$-divergence, evaluated at the uncontaminated model parameters.

To introduce outlier contamination, we assume the error terms/responses and the random effects are drawn from the underlying mixture distributions \citep[widely adopted in the literature for studying robust estimation e.g.,][]{Basu1998,F2013} as follows
\begin{equation*}\label{contami}
\begin{split}
f_c(y_{ij}|\bb_i; \bth^{\ast})&=(1-\zeta_{1ij})\phi(y_{ij};\bx_{ij}^{\top}\bbe^{\ast}+\bz_{ij}^{\top}\bb_i,\sigma^{\ast2})+\zeta_{1ij}\delta(y_{ij}|\bb_i),\\
f_c(\bb_i;\bth^{\ast})&=(1-\zeta_{2i})\phi_q(\bb_i;\zero,\bR^{\ast})+\zeta_{2i}\delta(\bb_i),
\end{split}
\end{equation*} 
where $\zeta_{1ij}$ and $\zeta_{2i}$ are contamination probabilities at the error term and random effects levels of the LMM, respectively, and $\delta(y_{ij}|\bb_i)$ and $\delta(\bb_i)$ are the corresponding contamination distributions. The above formulation leads to contaminated marginal distribution of the response for the $i$th cluster as 
$
f_c(\by_i;\bth^{\ast})=\int f_c(\bb_i;\bth^{\ast})\prod_{j=1}^{n_i}f_c(y_{ij}|\bb_i; \bth^{\ast})d\bb_i.
$

Under contamination, the robust estimator $\bthh_{\gamma}$ may not converge to $\bth_{\gamma}^{\ast}$ (or $\bth^{\ast}$), but instead to $\bth_{\gamma}^{\dagger}$, which is defined as the maximizer of $\lim_{m\to \infty}m^{-1}D_{\gamma}^{M\dagger}(\bth)$ where
\begin{equation*}
\begin{split}
\frac{1}{m} D_{\gamma}^{M\dagger}(\bth) 
&= \frac{\bn}{\gamma}\log\left(\frac1{m\bn}\sum_{i=1}^m\sum_{j=1}^{n_i}\int\phi(y_{ij};\bx_{ij}^{\top}\bbe+\bz_{ij}^{\top}\bbt_i(\by_i;\bth),\sigma^2)^{\gamma}f_c(\by_i;\bth^{\ast})dy_{ij}\right) \\
&
+\frac{1}\gamma\log\left(\frac1m\sum_{i=1}^m\int\phi_q(\bbt_i(\by_i;\bth);\zero,\bR)^{\gamma}f_c(\by_i;\bth^{\ast})dy_i\right) + C_{\gamma}(\bth).
\end{split}
\end{equation*} 
To clarify, $\lim_{m\to \infty}m^{-1}D_{\gamma}^{M\dagger}(\bth)$ is the limiting hierarchical $\gamma$-divergence evaluated at the contaminated model, as opposed to $\lim_{m\to \infty}m^{-1}D_{\gamma}^{M\ast}(\bth)$ which is evaluated based on the uncontaminated model. To establish robustness, we seek to demonstrate $\bth_{\gamma}^{\dagger}$ is reasonably close to the pseudo-true parameter $\bth_{\gamma}^{\ast}$ even under contamination.

To study robustness, we can introduce the following three types of contamination scenarios. Given $s_i=(s_{i1},\ldots,s_{in_i})^\top$ is an element of $D_i= \{0,1\}^{n_i}\setminus \{0\}^{n_i}$, which is used to indicate which mixture component the $y_{ij}$'s are generated from in the mixture distribution $f_c(y_{ij}|\bb_i; \bth^{\ast})$ for the $i$-th cluster, we have:
\begin{enumerate}[(i)]
    \item $h_1(\by_i;\bth)=\int \prod_{j=1}^{n_i} \phi(y_{ij};\bx_{ij}^{\top}\bbe+\bz_{ij}^{\top}\bb_i,\sigma^2)\delta(\bb_i)d\bb_i$. This is the marginal distribution of the response when the random effects are generated from their corresponding contamination distribution, but all errors are generated from their corresponding uncontaminated distribution.
    
    \item $h_2(\by_i;\bth) = \int \prod_{j=1}^{n_i}((1-\zeta_{1ij})\phi(y_{ij};\bx_{ij}^{\top}\bbe+\bz_{ij}^{\top}\bb_i,\sigma^2))^{1-s_{ij}}(\zeta_{1ij}\delta(y_{ij}|\bb_i))^{s_{ij}}\delta(\bb_i)d\bb_i$:  This is the marginal distribution when the random effects are generated from their corresponding contaminated distribution, and one or more of the errors are also generated from their corresponding contamination distribution.

\item $h_3(\by_i;\bth) = \int \prod_{j=1}^{n_i}((1-\zeta_{1ij})\phi(y_{ij};\bx_{ij}^{\top}\bbe+\bz_{ij}^{\top}\bb_i,\sigma^2))^{1-s_{ij}}
(\zeta_{1ij}\delta(y_{ij}|\bb_i))^{s_{ij}}\phi(\bb_i;\zero,\bR)d\bb_i$: This is the marginal distribution when the random effects are generated from their corresponding uncontaminated distribution, but one or more of the errors are generated from their corresponding contamination distribution. 
\end{enumerate}

Finally, we define the following two functions of $\bth$ that can be effectively interpreted as a measure of separation between the uncontaminated and contaminated distributions. 
\begin{align*}
&\nu_{1ik}(\bth)\equiv \int\sum_{j=1}^{n_i}\phi(y_{ij};\bx_{ij}^{\top}\bbe+\bz_{ij}^{\top}\bbt_i(\by_i;\bth),\sigma^2)^{\gamma}h_{k}(\by_i;\bth^{\ast})d\by_i\\
&\nu_{2ik}(\bth)\equiv \int\phi_q(\bbt_i(\by_i;\bth);\zero,\bR)^{\gamma}h_k(\by_i;\bth^{\ast})d\by_i.
\end{align*}
Both quantities depend on the contamination scenario. For $\nu_{1ik}(\bth)$, for a given cluster, the density $\phi(y_{ij};\bx_{ij}^{\top}\bbe+\bz_{ij}^{\top}\bbt_i(\by_i;\bth),\sigma^2)$ evaluated at the true parameters $\bth^{\ast}$ and the density $h_{k}(\by_i;\bth^\ast)$ should be well-separated from each other, provided the uncontaminated distribution $G(\by_i;\bth^\ast)$ and contamination distribution $h_{k}(\by_i;\bth^{\ast})$ are themselves well-separated. Hence we expect that, so long as $\gamma>0$ around the true parameter, then $\nu_{1ijk}(\bth)$ should be sufficiently small. A similar interpretation can be applied to $\nu_{2ik}(\bth)$. To formalize this, we can define the restricted parameter space 
$$
\Omega_{\nu}=\{\bth: \nu_{1ijk}(\bth)\leq \nu, \  \nu_{2ik}(\bth)\leq \nu, \  s_i\in D_i,\ i=1,\ldots,m,\  j=1,\ldots,n_i,\  k=1,2,3\}.
$$
Such a space is necessary to evaluate the latent bias of the estimator based on $\gamma$-divergence or weighed estimating equations \citep[see also][]{FE2008,F2013}. 
Finally, we assume the contamination probabilities $\zeta_{1ij}$ and $\zeta_{2i}$ are random and generated from some unknown distributions. Moreover, $\zeta_i = 1-(1-\zeta_{2i})\prod_{j=1}^{n_i}(1-\zeta_{1ij})$ is the contamination probability in the $i$th cluster, which is independent across the $m$ clusters and $\zeta_1,\ldots,\zeta_m$ have the same expectation (see the Supplementary Material for details). We then obtain the following result:

\begin{thm}\label{thm:robust}
For an arbitrary fixed $\nu$, suppose $\bth_{\gamma}^{\dagger}, \bth_{\gamma}^{\ast} \in \Omega_\nu$. 
Then under the regularity conditions given in the Supplementary Material, it holds that $\bth_{\gamma}^{\dagger}=\bth_{\gamma}^{\ast}+O(\nu)$.
\end{thm}

The difference expressed as $O(\nu)$ is known as ``latent bias".
To achieve robustness, it is desirable for $\nu$ to be small. 
However, assuming a small value of $\nu$ shrinks the parameter space $\Omega_\nu$ in which the pseudo-true parameters are assumed to exist, thus making the assumption stronger. Therefore, Theorem~\ref{thm:robust} represents a trade-off between the strength of robustness (i.e., the magnitude of latent bias) and the strength of the assumptions regarding the existence of the pseudo-true parameters. 
In other words,
when the uncontaminated and contaminated distributions are well-separated, i.e., $\nu$ is small, the pseudo-true parameter under no contamination is close to the pseudo-true parameter under contamination., thus demonstrating the robustness of the hierarchical $\gamma$-divergence for LMMs i.e., 
$\widehat{\bth}_\gamma=\bth_{\gamma}^{\ast}+o_p(1)+O(\nu)$ as $m\to\infty$ under contamination.

\section{Simulation Study}\label{sec:sim}

\subsection{Simulation design}
We conducted a simulation study to assess the estimation and inferential performance of the hierarchical $\gamma$-divergence estimator for LMMs, relative to maximum likelihood and several existing robust procedures. Specifically, we compared the following five methods: 
1) the proposed hierarchical $\gamma$-divergence estimator with tuning parameter either chosen using the H-score discussed in Section \ref{sec:selection}, or simply set to $\gamma=0.5$. For the former, we searched for the optimal $\gamma$ in the candidate set $\Gamma = \{0, 0.05,\ldots, 0.45, 0.50\}$.
We refer to these two approaches as aHGD (`a' for adaptive) and HGD (where $\gamma = 0.5$ is fixed), respectively; 
2) the standard Gaussian maximum likelihood estimator as implemented in \verb+lme4+ \citep{Bates2015}, which we refer to as ML; 
3) the robust estimator using Huber's $\psi$ function to robustify the estimating equations, as implemented in the \texttt{R} package \verb+robustlmm+ \citep{Koller2016}. We refer to this method as RML; 
4) the maximum likelihood estimator where the normality assumption is replaced with a $t$-distribution with four degrees of freedom in both the random effects and the error terms. This is an alternative approach to robustifying the LMM, and is implemented in the \texttt{R} package \verb+heavy+ \citep{heavy}. We refer to this method as HT; 
5) the robust estimator applying the density power divergence to the marginal likelihood as proposed by \cite{saraceno2023robust}. The  tuning parameter of the density power divergence is set to the recommended default value of $\alpha=0.5$. We refer to this method as mDPD.

We generate data from a (potentially) contaminated linear mixed model as follows. We set the number of clusters to $m=50,100$, and construct the clusters such that the first 20\% of the clusters have cluster size equal to $n_i = 10$, the second 20\% have cluster size equal to $n_i = 15$, and so on until the final 20\% of the clusters have a cluster size of $n_i = 30$. Next, we generate the response as
$$
y_{ij}=\beta_0 + \beta_1x_{1ij} + \beta_2x_{2ij} + \beta_3x_{3ij} + b_{0i} + b_{1i}x_{2ij}+\ep_{ij}, 
$$
where the vector of covariates $(x_{1ij}, x_{2ij}, x_{3ij})^{\top}$ is generated independently from a trivariate normal distribution with mean vector zero and a covariance matrix with ones along the diagonal and 0.4 for all off-diagonal elements. 
We set the vector of true fixed effect coefficients to $\bbe = (\beta_0,\beta_1,\beta_2,\beta_3)^{\top} = (0.5,0.3,0.5,0.8)^{\top}$, while for the error term $\ep_{ij}$ we adopted the contaminated distribution $\ep_{ij}\sim (1-\zeta_{1ij})N(0,\sigma^2)+\zeta_{1ij} N(a,1)$, with $\sigma=1.5$. 
The contamination probabilities (at the level of the error terms) are allowed to depend upon the covariate values themselves. In particular, we set $\zeta_{1ij}=2c_1/(1+\exp(3-x_{1ij}))$ so that error terms with larger $x_{1ij}$ are more likely to be outliers. 

Turning to the random effects $\bb_i=(b_{0i}, b_{1i})^\top$, we adopted the contaminated distribution $
\bb_i\sim (1-\zeta_{2i})N((0,0), \bR) + \zeta_{2i} N((a,a), {\rm Diag}(1,1))
$, where $\zeta_{2i}=c_2$ and $\bR$ is a $2 \times 2$ random effects covariance matrix with ones on the diagonals and 0.3 for all off-diagonal elements. 
To summarize, contamination can occur at both levels of the LMM, producing outliers in both the error terms $\ep_{ij}$ and the random effects $\bb_i$; the latter causes all the observations in the $i$-th cluster to be outlying observations. We set $a=10$ throughout this simulation study.

The quantities $c_1$ and $c_2$ control the contamination probabilities for the error components and random effects, respectively. 
We considered the following nine scenarios for the degree of contamination:
\begin{align*}
\begin{array}{lllll}
({\rm S}1): (c_1, c_2)=(0,0), \ \ \ &({\rm S}2): (c_1, c_2)=(0, 0.05), \ \ \ &({\rm S}3): (c_1, c_2)=(0, 0.1),\\
({\rm S}4): (c_1, c_2)=(0.05, 0), &({\rm S}5): (c_1, c_2)=(0.05,0.05), &({\rm S}6): (c_1, c_2)=(0.05,0.1),\\
({\rm S}7): (c_1, c_2)=(0.1, 0), &({\rm S}8): (c_1, c_2)=(0.1, 0.05), &({\rm S}9): (c_1, c_2)=(0.1,0.1).
\end{array}
\end{align*}
There is no contamination in Scenario 1, so we expect standard maximum likelihood estimation (ML) to work well here. 
In the other eight settings, contamination occurs in either the random effects and/or observation process, consistent with the three types of contamination scenarios defined in Section \ref{sec:robust}. Scenario 9 is the most extreme, where there are non-negligible numbers of clusters and observations coming from their respective contamination distributions. For each scenario, we simulated 500 datasets.

To assess performance, we computed the empirical mean squared errors (MSE), averaged across the 500 simulated datasets, of the estimators of the fixed regression coefficient vector $\bbe$, the error variance $\sigma^2$, the random effects covariance matrix $\bR$, and the predictions of the random effects $\bb_i$. Note the average MSE of $\bbe$ and $\bR$ is computed across all their respective elements, while for the random effects it is computed over all the clusters and elements. Also, because the scale parameters in the HT method are not equal to the variance parameters in the true model, we do not include this method for comparison when assessing performance on estimating $\bR$ and $\sigma$. 
Focusing on the fixed effect coefficients $\bbe$, we also assessed the performance of confidence intervals constructed using the clustered bootstrap method in Section \ref{sec:inference}, where to ease the computational burden of this part of the simulation study, we only considered 200 simulated datasets. 
For each simulated dataset, we obtained $100$ bootstrap estimates for the aHGD and HGD estimators and computed 95\% percentile confidence intervals for the vector of fixed effect coefficients $\bbe$. We then calculated the empirical coverage probability, along with the interval score \citep{gneiting2007strictly} to assess performance. 
For the $k$-th fixed effect coefficient, the latter is computed as 
$
{\rm IS}_k=(1/200)\sum_{r=1}^{200}({\rm CI}_{k(r)}^u - {\rm CI}_{k(r)}^l+(2/0.05)((\beta_k-{\rm CI}_{k(r)}^u)_+ + ({\rm CI}_{k(r)}^l-\beta_k)_+)),
$
where ${\rm CI}_{k(r)}^u$ and ${\rm CI}_{k(r)}^l$ denotes the upper and lower values of the confidence interval for $\beta_k$ in the simulated dataset $r$, and $(x)_{+}=\max(0, x)$.
We compared the above intervals from aHGD and HGD with bootstrap confidence intervals for the ML and RML estimators, computed in the same way as aHGD and HGD. Note however that the computation time of RML is much longer than the other methods including HGD e.g. with $m=50$, RML takes several minutes while HGD takes less than ten seconds. Therefore, to ensure the simulation study remained feasible we opted to relaxed the convergence criteria in \verb+robustlmm+ package to make the computation time similar to that of HGD.  
Finally, we also considered computed Wald-type confidence intervals for ML and RML based on standard errors obtained as part of their software,
but found the performance of these to be very similar to those their bootstrap-based counterparts, and so have omitted them from the presentation below.

\subsection{Results}
\begin{figure}[htb!]
\centering
\includegraphics[width=0.8\textwidth]{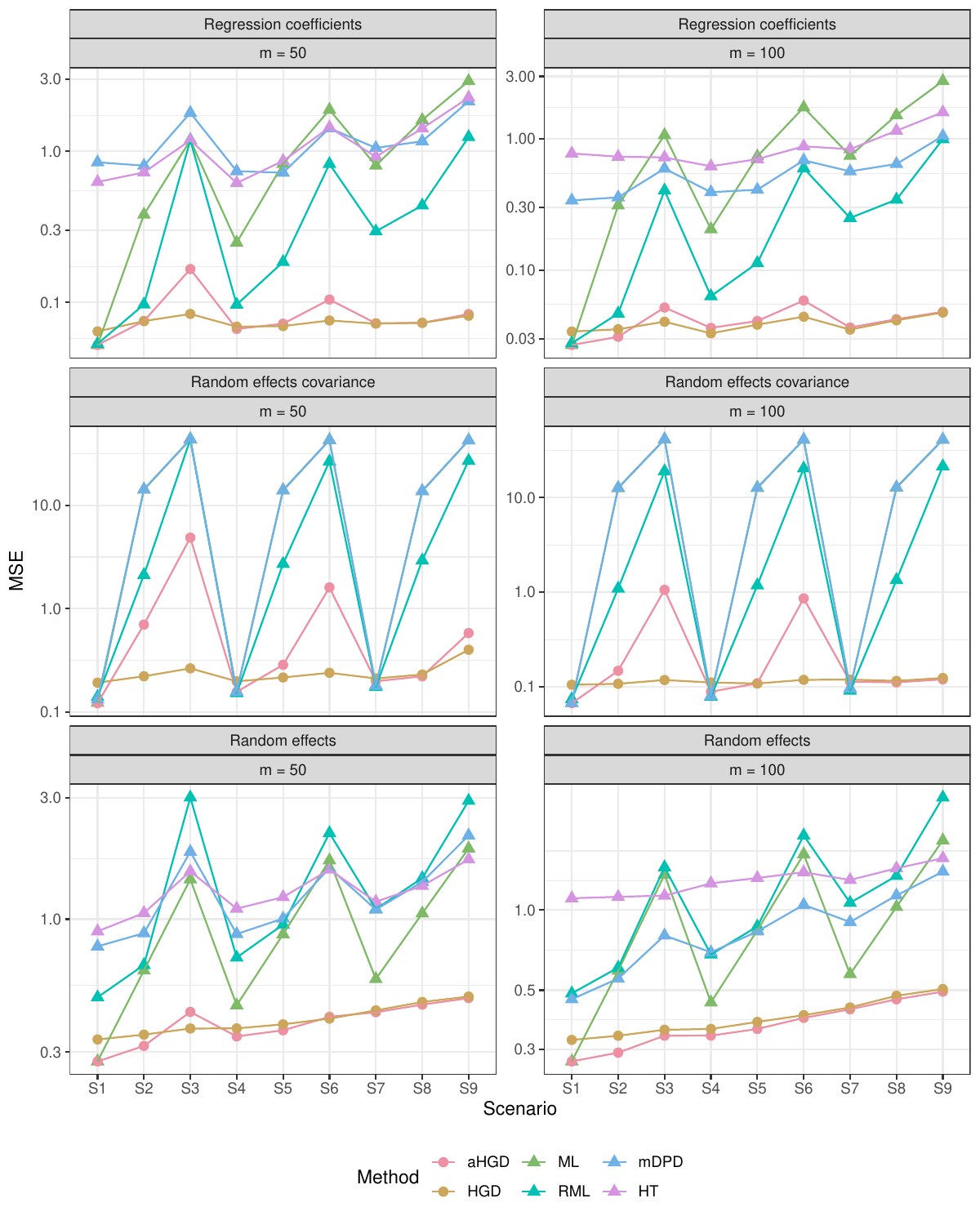}
\caption{Mean squared errors (MSE) for different estimators of the regression coefficients $\bbe$ (top row), random effect covariance matrix $\bR$ (middle row) and random effects $\bb_i$ (bottom row) across nine different contamination scenarios and $m=50$ (left) and $m=100$ (right) clusters in the LMM. 
\label{fig:sim-main}
}
\end{figure}

Figure~\ref{fig:sim-main} presents the MSE values for regression coefficients, the random effect covariance matrix, and random effects, while for brevity the results of the error variance are provided in the Supplementary Material. 
Overall, the proposed hierarchical $\gamma$-divergence estimator performed best in terms of point estimation/prediction, with often the lowest MSEs when comparing across the nine scenarios considered. 
The performance of all six methods for estimation of the $\bbe$ and $\sigma^2$ was relatively similar when there were no outliers in the error terms (scenarios 1, 2 and 3), while substantial differences emerged once there was a non-zero level of contamination especially in the error terms (scenarios 4-9). 
By contrast, when it came to estimation of the random effects covariance matrix $\bR$ and prediction of the effects $\bb_i$, the most prominent differences were observed when there was a non-zero level of contamination in the random effects (scenarios 2-3, 5-6, 8-9).

The aHGD estimator (with adaptive $\gamma$) tended to perform better than HGD with fixed $\gamma=0.5$. 
Indeed, the average chosen values of $\gamma$ from aHGD across all the nine contamination settings considered was smaller than $0.5$ (Table~\ref{tab:gamma}). On the other hand, when the contamination level is higher at the random effects level e.g., $\zeta_1 = 0.1$ and scenarios 3, 6, and 9, HGD tended to outperform aHGD especially in the estimation of the random effect covariance matrix $\bR$, suggesting that tuning $\gamma$ using the H-score was somewhat conservative in robustifying the LMM. Turning to the other estimators, the standard (non-robust) ML estimator tended to perform worst when there was contamination in the error terms. The HT approach using the $t$-distribution tended to perform better than standard ML for estimating the model parameters, but its performance was still poor relative to the other three robust approaches. In scenarios 7-9 where there was strong contamination in both the random effects and error terms, both the proposed aHGD and HGD estimators strongly outperformed the competitors.

\begin{table}[tb]
\caption{Average values of chosen $\gamma$ in the  hierarchical $\gamma$-divergence (aHGD) estimator across nine different contamination scenarios and $m=50$ and $m=100$ clusters in the LMM, where $\gamma$ was tuned using the H-score.
\label{tab:gamma}
}
\begin{center}
{\small
\begin{tabular}{cccccccccccc}
\toprule[1.5pt]
Scenario &  & s1 & s2 & s3 & s4 & s5 & s6 & s7 & s8 & s9 \\
\cmidrule{3-11}
$m=50$ &  & 0.002 & 0.143 & 0.238 & 0.213 & 0.219 & 0.268 & 0.277 & 0.279 & 0.300 \\
$m=100$ &  & 0.000 & 0.130 & 0.234 & 0.214 & 0.217 & 0.272 & 0.278 & 0.282 & 0.292 \\
\bottomrule[1.5pt]
\end{tabular}
}
\end{center}
\end{table}

\begin{figure}[htb!]
\centering
\includegraphics[width=0.95\textwidth]{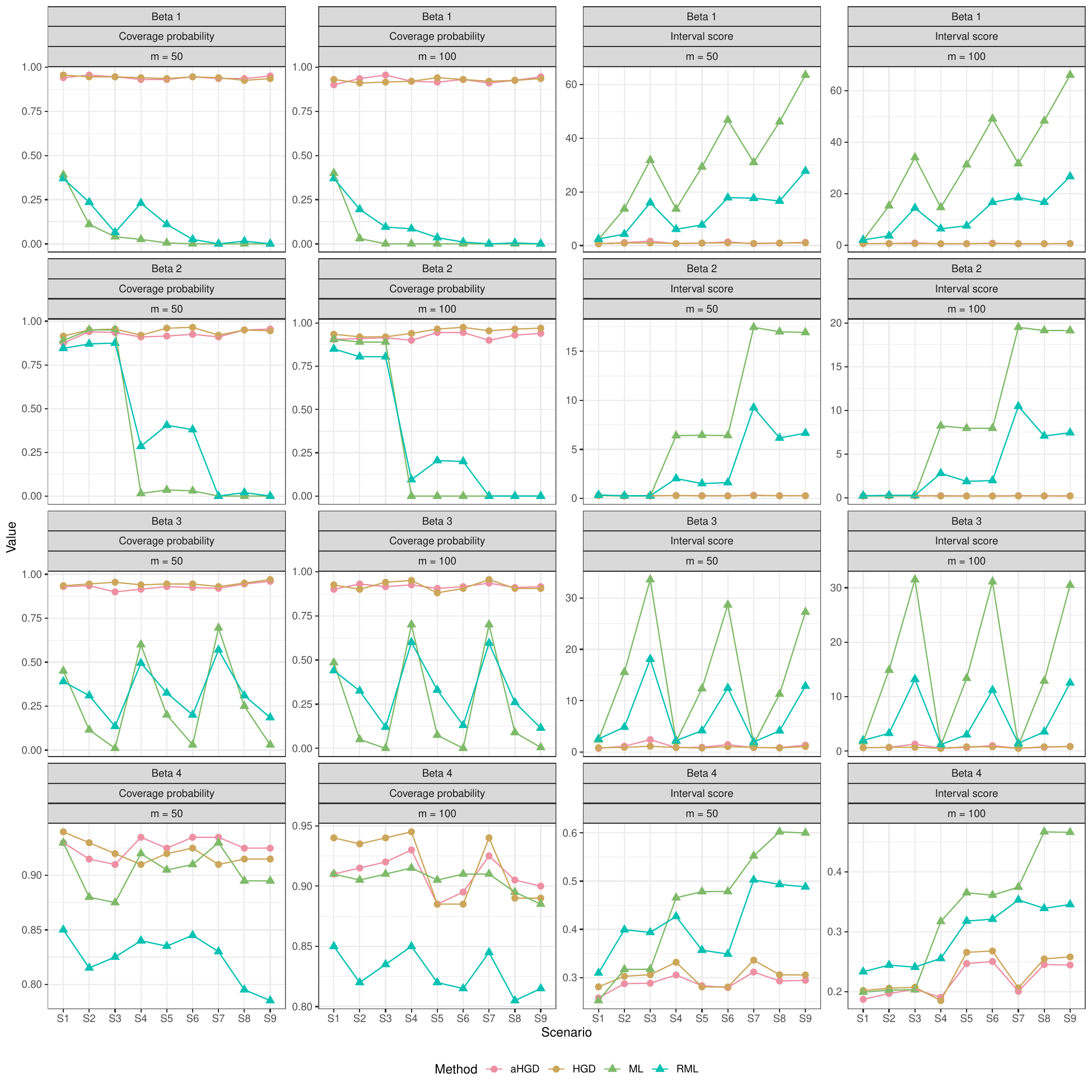}
\caption{
Empirical coverage probabilities (first and second columns) and interval scores (third and fourth columns) for $95\%$ bootstrap-based confidence intervals of the four fixed effects coefficients (top to bottom), across the nine different contamination scenarios and $m=50$ and $m=100$ clusters in the LMM.
}
\label{fig:sim-CI}
\end{figure}

Finally, Figure \ref{fig:sim-CI} presents results for the confidence intervals of the fixed effects coefficients. Despite the relatively low number of bootstrap samples, the empirical coverage probability of both hierarchical $\gamma$-divergence estimators was close to the nominal $95\%$ level and their interval scores were reasonable across the four coefficients and all nine scenarios considered. By contrast, the performance both of ML and (perhaps surprisingly) RML was much lower than the nominal level when the data were contaminated at either the error term and/or random effects level. 
Finally, in the Supplementary Material, we present an additional simulation scenario where the data generation processes was set up to better mimic the characteristics of the error and random effects distributions observed in the application in Section~\ref{sec:app} e.g., evidence of skewness in the random effects. Overall, these results also indicate robustness of GHD. 
  
\section{Application to AIDS Cohort Study}\label{sec:app}
We illustrate an application of the hierarchical $\gamma$-divergence estimator for LMMs on data from the Multi-center AIDS Cohort Study, available in the \texttt{R} package \texttt{catdata} \citep{catdata}. 
The dataset originates from a survey of $m=369$ men infected with HIV (human immunodeficiency virus). As the response, the number of CD4 cells was measured for each person over time, with the number of measurements $n_i$ ranging from 1 to 12.  The response variable CD4 cell count was bounded away from zero and so can be treated as a continuous variable. As covariates, information was also collected on years since seroconversion (Time), an indicator of recreational drug use (Drug), the number of sexual partners (Partners), the number of packs of cigarettes smoked each day (Packs), a mental illness score (Cesd), and baseline age centered around 30 (Age).
We standardized all covariates to have mean zero and variance equal to one.  
Following \cite{ZD1994} among others, who analyzed the dataset with semiparametric components for Time, Cesd and Age, we also included quadratic and cubic terms for these three continuous covariates. The results in $p = 13$ predictors in total including the intercept.

\begin{figure}[tb]
\centering
\includegraphics[width=\textwidth,clip]{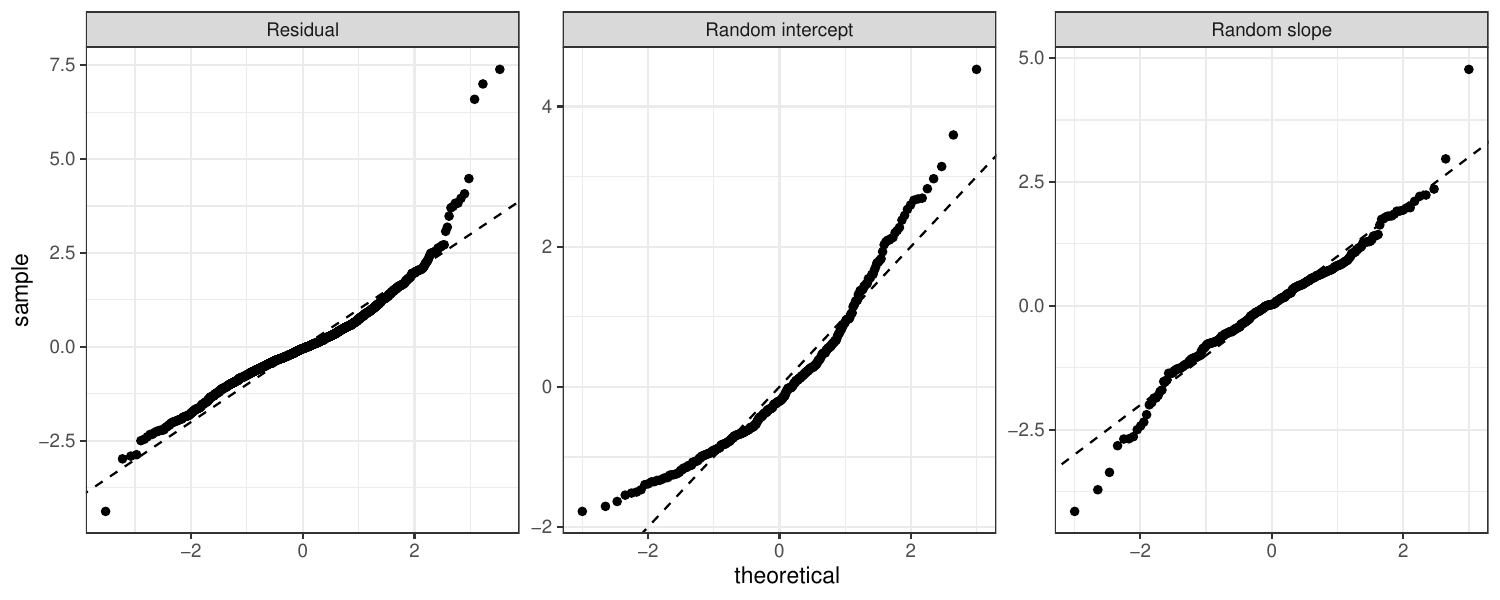}
\caption{Normal probability plots of the standardized residuals (left), random intercept (center) and random slope (right), based on fitting the LMM to the AIDS cohort study using standard maximum likelihood estimation.
\label{fig:resid}
}
\end{figure}

We fitted the LMM in equation \eqref{LMM}, where $\bx_{ij}$ is a vector of the 13 predictors discussed above, and $\bz_{ij}$ involves a random intercept and linear slope for time.
We began by using standard maximum likelihood estimation, and computed the standardized residuals $\sih^{-1}(y_{ij}-\bx_{ij}^{\top}\beh-\bz_{ij}^{\top}\bbh_i)$. Figure~\ref{fig:resid}, presents a normal probability plot of these results, along with corresponding plots for the predicted random intercepts and slopes for time. All plots provide evidence that the normality assumption for the error-terms and random effects is violated, suggesting some potentially outlying positive errors, and potentially outlying clusters having both relatively large positive or negative random intercepts and slopes. 

\begin{figure}[htb]
\centering
\includegraphics[width=0.9\textwidth]{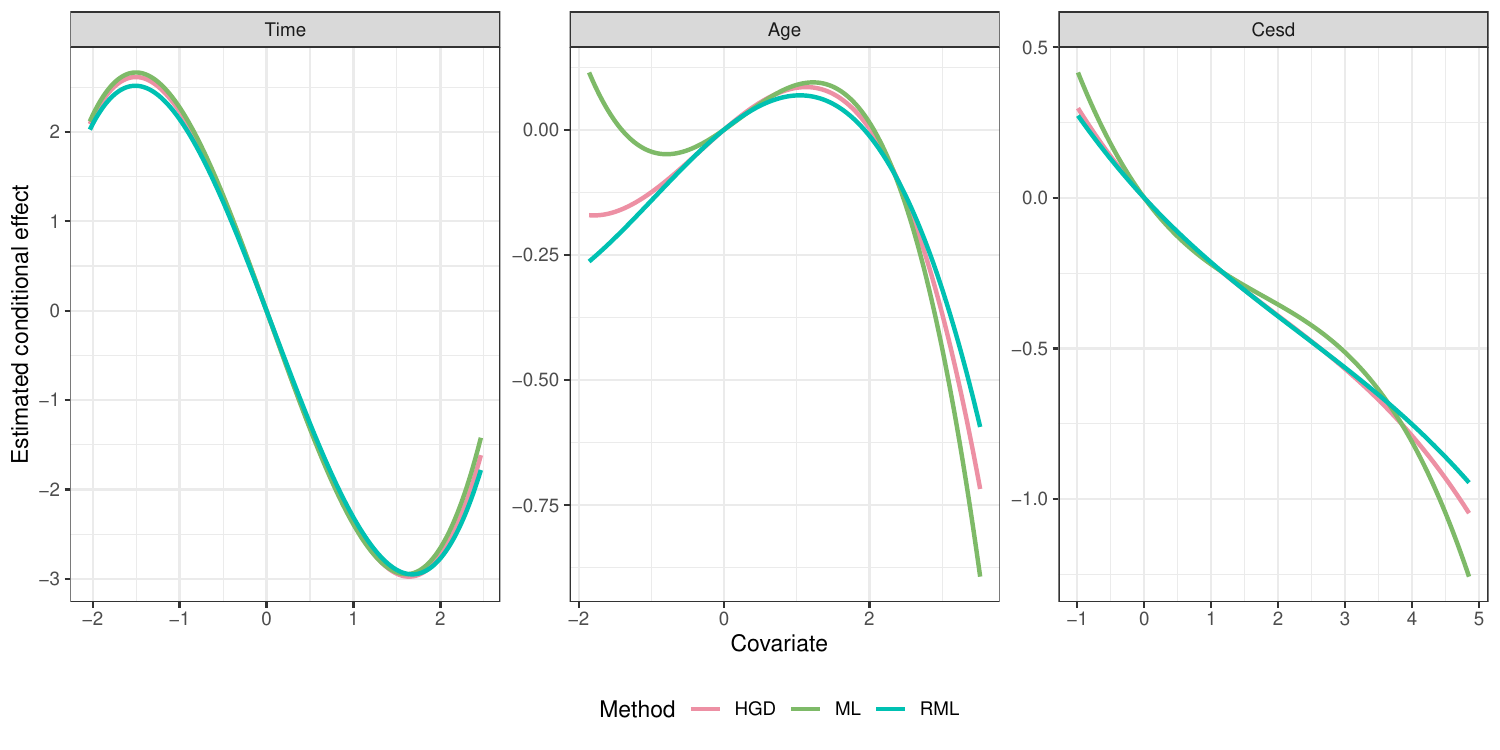}
\caption{Estimated conditional effects of three covariates (Time, Age, and Cesd) in the application of LMMs to the  AIDS cohort study.
\label{fig:effect}
}
\end{figure}

With the above findings in mind, we proceeded to use the hierarchical $\gamma$-divergence estimator to fit a robust LMM, comparing the results with the standard maximum likelihood and robust maximum likelihood (RML) estimator of \citet{Koller2016} as used in the simulation study in Section \ref{sec:sim}. 
We tuned the parameter $\gamma$ using H-scores as detailed in Section~\ref{sec:selection}, where we selected $\gamma$ from the range $\Gamma = \{0, 0.01,\ldots,0.19, 0.20\}$. This resulted in $\gamma=0.08$ being selected.  Figure~\ref{fig:effect} presents conditional plots of the non-linear effects of Time, Age, and Cesd based on the three estimators, holding all other variables at their mean and setting the random effects to zero. All three approaches offer comparable estimated effects of Time, but differ markedly especially in the tails for Age and Cesd. For Age, the ML estimator produces a strong and potentially implausible upward trend at lower ages; this may be due to the presence of outlying observations as suggested in Figure \ref{fig:resid}. By contrast, the RML estimator produces a more or less quadratic effect for Age, while the hierarchical $\gamma$-estimator lies somewhere between the ML and RML estimators, with a levelling out of the effect at lower ages. The results from the hierarchical $\gamma$-estimator also to a large extent lie in between the ML and RML estimators when we examine the conditional effects for Time and Cesd.

Table~\ref{tab:est} presents point estimates along with confidence intervals (constructed using the same approach discussed in the simulations in Section \ref{sec:sim}) for all three methods. All three approaches offer the same conclusions for the effect of Drugs (with no statistically clear evidence of an effect), as well as the effect of Partners (evidence that more sexual partners is associated with a higher CD4 cell count). The effect of Packs, Time, and Cesd was less extreme (but still statistically significant) for the hierarchical $\gamma$-divergence compared to ML and RML, showing the strong robustness properties of HGD.

\begin{table}[tb]
\caption{Point estimates and their $95\%$ confidence intervals (CI) of regression coefficients of the three methods used to fit LMMs to the AIDS cohort study. Covariate effects whose corresponding 95\% confidence interval exclude zero are highlighted with an asterisk.
\label{tab:est}
}
\begin{center}
{
\begin{tabular}{ccccc}
\toprule[1.5pt]
Covariate && ML & RML & HGD  \\
\cmidrule{3-5}
Drugs &  & 0.06 (-0.09, 0.21) & 0.10 (-0.03, 0.23) & 0.07 (-0.09, 0.22) \\
Partners &  & 0.17 (0.01, 0.32)* & 0.18 (0.05, 0.31)* & 0.15 (0.01, 0.29)* \\
Packs &  & 0.37 (0.17, 0.57)* & 0.31 (0.13, 0.48)* & 0.37 (0.15, 0.60)* \\
Time &  & -2.71 (-2.98, -2.44)* & -2.56 (-2.8, -2.33)* & -2.66 (-2.98, -2.38)* \\
Time$^2$ &  & -0.06 (-0.18, 0.06) & -0.09 (-0.19, 0.02) & -0.07 (-0.18, 0.04) \\
Time$^3$ &  & 0.37 (0.29, 0.46)* & 0.34 (0.27, 0.41)* & 0.36 (0.28, 0.45)* \\
Cesd &  & -0.30 (-0.51, -0.09)* & -0.24 (-0.42, -0.06)* & -0.25 (-0.44, -0.06)* \\
Cesd$^2$ &  & 0.02 (-0.27, 0.32) & -0.04 (-0.31, 0.24) & -0.02 (-0.28, 0.27) \\
Cesd$^3$ &  & -0.04 (-0.17, 0.10) & -0.01 (-0.14, 0.12) & -0.02 (-0.17, 0.09) \\
Age &  & 0.10 (-0.34, 0.55) & 0.12 (-0.3, 0.53) & 0.12 (-0.27, 0.61) \\
Age$^2$ &  & 0.10 (-0.10, 0.30) & 0.03 (-0.14, 0.20) & 0.04 (-0.13, 0.23) \\
Age$^3$ &  & -0.02 (-0.07, 0.03) & 0.00 (-0.05, 0.04) & -0.01 (-0.05, 0.03) \\
\bottomrule[1.5pt]
\end{tabular}
}
\end{center}
\end{table}

We report the estimated error variance and covariance matrix of the random effects in Table~\ref{tab:est-variance}.
The HGD estimates are quite different from those from (non-robust) ML, possibly due to the existence of outliers, but RML provides even more different estimates than ML. 
This could be because the contamination is not so heavy in this application (as the optimal $\gamma$ in HGD is $0.08$) that RML might be more robust than necessary. 
Moreover, scatter plots of the predicted random effects (see Supplementary Material) suggest that the predictors from HGD are typically more shrunk toward the origin compared  with the others, including RML, reflecting the idea that HGD can successfully protect random effect prediction from outliers. 

Finally, in the Supplementary Material at the suggestion of one of the reviewers, we report results based on applying one of three potential three transformation variance-stabilizing functions (logarithmic, square root and cube root transformations) to the response and repeating all of the above analyses for each. Overall, results suggest there still exist outliers even after such data transformations, suggesting the existence of outliers as observed in Figure~\ref{fig:resid} cannot be simply solved by data transformation. Moreover, the general conclusions reached regarding the statistical significance of the covariates are broadly similar across all three transformations, and are similar to those of the original untransformed analysis presented above.

\begin{table}[tb]
\caption{Estimates of the error variance $\sigma^2$ and covariance matrix $R$ under ML and HGD methods. 
\label{tab:est-variance}
}
\begin{center}
\begin{tabular}{ccccccc}
\toprule[1.5pt]
 &  & $\sigma^2$ & $R_{11}$ & $R_{12}$ & $R_{22}$ \\
 \cmidrule{3-6}
ML &  & 5.20 & 5.75 & -0.55 & 1.60 \\
RML &  & 3.41 & 4.97 & -0.35 & 1.45 \\
HGD &  & 4.63 & 5.45 & -0.40 & 1.87 \\
\bottomrule[1.5pt]
\end{tabular}
\end{center}
\end{table}

\section{Discussion}\label{sec:dis}
We proposed a new hierarchical $\gamma$-divergence estimator for robust estimation and inference in linear mixed models. The method is based on robustifying the components of the likelihood associated with both the response/error and the random effects in the LMM. We developed a computationally efficient iterative procedure for computing the parameter estimates and random effect predictions, associating the hierarchical $\gamma$-divergence with both the MM-algorithm and a series of weighted estimated equations with normalized powered density weights. Given the form of the weighted estimating equations, we also proposed a clustered bootstrap approach for uncertainty quantification. Simulation studies demonstrate the strong performance of the proposed method compared to existing robust methods in terms of handling severe outlier contamination (in the errors and/or random effects) and statistical efficiency. An application to data from a AIDS cohort study highlights the importance of employing robust methods in fitting LMMs in the presence of potential outliers.

Viewing hierarchical $\gamma$-divergence as fitting a LMM using a set of weighted estimating equations with normalized density weights, suggests that one potential avenue of future research is replacing these weights (and thus the $\gamma$-divergence measure) with alternative weights such as those derived from a density power divergence \citep{Basu1998}. However, the use of other divergence measures does not necessarily ease optimization e.g., the ascent property of the MM algorithm may no longer hold, and it is unclear what theoretical robustness properties could be derived in these other possibilities. Another avenue of future research is to extend the hierarchical $\gamma$-divergence to multilevel LMMs, potentially with different tuning parameters $\gamma$ at difference levels of the hierarchy, and subsequently generalized linear mixed models. An immediate challenge with the latter is that in the non-normal response case, the marginal log-likelihood function, and thus attempts to replace it with an alternative robust measure, does not possess a closed-form. One approach to overcome this maybe to work with robustifying a quasi- or variational likelihood function instead \citep{breslow1993approximate,ormerod2012gaussian,Hui2017}.
Yet another related possibility is to robustify a version of the restricted maximum likelihood (REML) estimator using the $\gamma$-divergence framework for the purposes of improved variance component estimation, although one potential drawback may be that estimates of the fixed effect coefficients and random effect predictions may be obtained via a separate (robustified) objective function.  
Regarding the computation algorithm, although the MM algorithm is employed in the proposed HGD method, it would be worthwhile to explore alternative approaches, such as the Expectation-Maximization (EM) algorithm \citep{dempster1977maximum}, which is commonly used for estimating LMMs.
Finally, in practice the number of fixed and random effect covariates may be large, in which case augmenting the hierarchical $\gamma$-divergence with regularization penalties or information criterion for joint fixed and random selection effects may be appropriate \citep[see][and references therein]{lin2013fixed,hui2021use}.

\section*{Acknowledgement}
We would like to thank two anonymous reviewers and an associate editor for their valuable comments and suggestions. 
We also thank Robert Clark and Takeshi Kurosawa for useful discussions. 
SS was partially supported by Japan Society for Promotion of Science (KAKENHI) grant numbers 18K12757 and 21H00699. 
FKCH and AHW were supported by Australian Research Council Discovery Grants DP230101908.
The authors report there are no competing interests to declare.

\bibliographystyle{apalike}
\bibliography{refs}

\newpage
\begin{center}
{\LARGE {\bf Supplementary Material for ``Robust Linear Mixed Models using Hierarchical Gamma-Divergence"}}
\end{center}

\setcounter{equation}{0}
\renewcommand{\theequation}{S\arabic{equation}}
\setcounter{section}{0}
\renewcommand{\thesection}{S\arabic{section}}
\setcounter{lem}{0}
\renewcommand{\thelem}{S\arabic{lem}}
\setcounter{table}{0}
\renewcommand{\thetable}{S\arabic{table}}
\setcounter{figure}{0}
\renewcommand{\thefigure}{S\arabic{figure}}
\setcounter{page}{1}

\vspace{5mm}
This supplementary material provides some technical details, the proof of Theorem~1 and additional simulation results.

\section{Derivation of the modified joint log-likelihood}
It suffices to show that $\max_b L_J(\bth, \bb)=L_M(\bth)$, where $L_J(\bth, \bb)$ and $L_M(\bth)$ are defined in (\ref{JL}) and (\ref{mL}), respectively. 
We first note that $L_J(\bth, \bb)=\sum_{i=1}^m L_{J,i}(\bb_i,\bth)$, where 
\begin{equation*}
\begin{split}
L_{J,i}(\bb_i,\bth)= &
\frac{n_i}{2}\log(\sigma^2) -\frac12\log\det(\bSi_i)-\frac{1}{2\sigma^2}(\by_i-\bX_i\bbe-\bZ_i\bb_i)^\top (\by_i-\bX_i\bbe-\bZ_i\bb_i)\\
& \ \ \ \  
+\frac12\log\det(\bR)-\frac12 \bb_i^\top \bR^{-1}\bb_i,
\end{split}
\end{equation*}
and constants with respect to $\bth$ and $\bb$ are ignored. Then, $\bb_1,\ldots,\bb_m$ can be separately maximized given $\bth$ and the maximizer of $\bb_i$ is $\bbt_i(\bth)=\bR\bZ_i^\top \bSi_i^{-1}(\by_i-\bX_i\bbe)$. Hence, we have 
\begin{equation*}
\begin{split}
L_{j,i}(\bbt_i(\bth),\bth)&=
-\frac12\log\det(\bSi_i)-\frac12(\by_i-\bX_i\bbe)^\top \bSi_i^{-1}(\by_i-\bX_i\bbe)\\
&\propto \phi_{n_i}(\by_i;\bX_i\bbe,\bSi_i).
\end{split}
\end{equation*}

\section{Derivation of the hierarchical $\gamma$-divergence}

Consider first the $\gamma$-divergence of the conditional distribution $\phi(y_{ij}; \bx_{ij}^\top \bbe+\bz_{ij}^\top \bb_i, \sigma^2)$. From straightforward calculations, we obtain
$$
\int \phi(y; \bx_{ij}^\top \bbe+\bz_{ij}^\top \bb_i, \sigma^2)^{1+\gamma}dy=(1+\gamma)^{-1/2}(2\pi\sigma^2)^{-\gamma/2}.
$$
This gives the following form of the $\gamma$-divergence:
\begin{equation}\label{sup-eq:gam1}
\frac{N}{\gamma}\log\left\{\frac{1}{N}\sum_{i=1}^m\sum_{j=1}^{n_i}\phi(y_{ij}; \bx_{ij}^\top \bbe+\bz_{ij}^\top \bb_i, \sigma^2)^\gamma\right\}+\frac{N\gamma}{2(1+\gamma)}\log\sigma^2 + C_1,
\end{equation}
where $C_1=N\log\{(2\pi)^{\gamma}(1+\gamma)\}/2(1+\gamma)$ is a constant independent of the parameters. 
Similarly, for the random effect distribution $\phi(\bb_i; \zero, \bR)$, it holds that 
$$
\int \phi(\bb; \zero, \bR)^{1+\gamma}d\bb=(1+\gamma)^{-q/2}\det(2\pi\bR)^{-\gamma/2}.
$$
This gives the following form of the $\gamma$-divergence:
\begin{equation}\label{sup-eq:gam2}
\frac{m}{\gamma}\log\left\{\frac{1}{m}\sum_{i=1}^m\phi(\bb_i; \zero, \bR)^\gamma\right\}+\frac{m\gamma}{2(1+\gamma)}\log \det(\bR) + C_2,
\end{equation}
where $C_2=mq\log\{(2\pi)^{\gamma}(1+\gamma)\}/2(1+\gamma)$.
Finally, by replacing the first and second terms in the modified joint likelihood (\ref{JL}) with (\ref{sup-eq:gam1}) and (\ref{sup-eq:gam2}), respectively, we obtain the objective function (\ref{Q}) as required.

\section{Derivation of the two H-scores }
Let $D_{\gamma}^{(1)}$ be the sum of the first and second terms of (\ref{Q}).
We then consider the transformed $\gamma$-divergence $\exp(D_{\gamma}^{(1)}/N)$, given by 
$$
\frac1N\sum_{i=1}^m\sum_{j=1}^{n_i}\phi(y_{ij};\bx_{ij}^\top \bbe + \bz_{ij}^\top \bb_i, \sigma^2)^{\gamma}/\gamma C_{\gamma}(\sigma^2), 
$$
where 
$C_{\gamma}(\sigma^2) = ((1+\gamma)^{-1/2}(2\pi\sigma^2)^{-\gamma/2})^{\gamma/(1+\gamma)}$.
The transformed $\gamma$-divergence can be regarded as the log-likelihood of the pseudo-model
$$
\log f(y_{ij}|\bb_i)\propto \phi(y_{ij};\bx_{ij}^{\top}\bbe+\bz_{ij}^{\top}\bb_i,\sigma^2)^{\gamma}/\gamma C_{\gamma}(\sigma^2).
$$
Following \cite{sugasawa2021selection}, one can define the H-score based on the pseudo-model to select $\gamma$. 
It should be noted that 
$$
\frac{\partial}{\partial y_{ij}}\log f(y_{ij}|\bb_i)
=-\frac{y_{ij}-\mu_{ij}}{\sigma^2 C_{\gamma}(\sigma^2)}\phi(y_{ij};\mu_{ij},\sigma^2)^{\gamma},
$$
and 
$$
\frac{\partial^2}{\partial y_{ij}^2}\log f(y_{ij}|\bb_i)=
-\frac{1}{\sigma^2 C_{\gamma}(\sigma^2)}\phi(y_{ij};\mu_{ij},\sigma^2)^{\gamma}
+\frac{\gamma(y_{ij}-\mu_{ij})^2}{\sigma^4 C_{\gamma}(\sigma^2)}\phi(y_{ij};\mu_{ij},\sigma^2)^{\gamma},
$$
where $\mu_{ij}=\bx_{ij}^{\top}\bbe+\bz_{ij}^{\top}\bb_i$.
These expressions lead to the form of $H_1(\gamma)$.

Similarly, we consider the transformed $\gamma$-divergence of the model $\bb_i\sim N(\zero, \bR)$, given by 
$$
\sum_{i=1}^m
\frac{\phi_q(\bb_i;\zero,\bR)^{\gamma}}{\gamma C_{\gamma}(\bR)},
$$
where $C_{\gamma}(\bR)$ is given in Section~\ref{sec:selection}.
This can be regarded as the log-likelihood of the pseudo-model
$$
\log f(\bb_i)\propto 
\frac{\phi_q(\bb_i;\zero,\bR)^{\gamma}}{\gamma C_{\gamma}(\bR)}.
$$
Then, it follows that 
$$
1_q^\top \frac{\partial}{\partial \bb_i}\log f(\bb_i)=
\frac{1_q \bR^{-1}\bb_i}{C_{\gamma}(\bR)}\phi(\bb_i;\zero,\bR)^{\gamma}
$$
and 
$$
{\rm tr}\left(\frac{\partial^2}{\partial \bb_i\partial \bb_i^\top}\log f(\bb_i)\right)
=\frac{\gamma(1_q \bR^{-1}\bb_i)^2-{\rm tr}(\bR^{-1})}{C_{\gamma}(\bR)}\phi(\bb_i;\zero,\bR)^{\gamma}.
$$

\section{Regularity Conditions and Proof of Theorem \ref{thm:robust}}

We assume the following standard regularity conditions:

\begin{itemize}
\item[(C1)]
There exists a constant $C_1$ such that $\max_{i=1,\ldots,m}n_i<C_1$.

\item[(C2)]
For every $i=1,\ldots,m$ and $j=1,\ldots,n_i$, there exists a sufficiently large constant $C_2$ such that $\|\bx_{ij}\|_{\infty}<C_2$ and $\|\bz_{ij}\|_{\infty}<C_2$, where $\|\cdot\|_{\infty}$ is the L$_\infty$-norm. 
Furthermore, the matrices $n_i^{-1}\sum_{j=1}^{n_i}\bx_{ij}\bx_{ij}^{\top}$ and $n_i^{-1}\sum_{j=1}^{n_i}\bz_{ij}\bz_{ij}^{\top}$ are positive definite with eigenvalues in the interval $[1/c_0, c_0]$, where $c_0$ is some large enough positive constant. 

\item[(C3)]
$\zeta_{1ij}$ and $\zeta_{2i}$ are mutually independent of $\by_i$ and $\bb_i$. 
Furthermore, the marginal probabilities $\zeta_i=1-\prod_{j=1}^{n_i}(1-\zeta_{1ij})(1-\zeta_{2i})$ are mutually independent and have the same mean $\bzeta$.

\item[(C4)]
The restricted paraemter space $\Omega_{\nu}$ is compact. 
\end{itemize}

\noindent
Now, we give the proof of Theorem~\ref{thm:robust}.
\begin{proof}
We can expand $f_c(\by_i;\bth)$ as 
\begin{align*}
f_c(\by_i;\bth)
&=\Big[\prod_{j=1}^{n_i}(1-\zeta_{1ij})\Big](1-\zeta_{2i})G(\by_i;\bth)
+\Big[\prod_{j=1}^{n_i}(1-\zeta_{1ij})\Big]\zeta_{2i}h_1(\by_i;\bth)\\
& \ \ \ \ 
+\zeta_{2i}\sum_{s_i\in D_i}h_2(\by_i;\bth)
+(1-\zeta_{2i})\sum_{s_i\in D_i}h_3(\by_i;\bth),
\end{align*}
where $h_1,h_2$ and $h_3$ are the three main contamination scenarios defined in the main text, and $G(\by_i;\bth)$ is the marginal distribution without contamination, i.e.,
$$
G(\by_i;\bth)=\int\phi(\bb_i;\zero, \bR)\prod_{j=1}^{n_i}\phi(y_{ij}; \bx_{ij}^\top\bbe+\bz_{ij}^\top \bb_i, \sigma^2)d\bb_i.
$$
Then, for $\bth\in \Omega_\nu$, it follows that 
\begin{align*}
&\int\phi(y_{ij};\bx_{ij}^{\top}\bbe+\bz_{ij}^{\top}\bbt_i(\by_i;\bth),\sigma^2)^{\gamma}f_c(\by_i;\bth^{\ast})dy_i \\
&=(1-\zeta_i)\int\phi(y_{ij};\bx_{ij}^{\top}\bbe+\bz_{ij}^{\top}\bbt_i(\by_i;\bth),\sigma^2)^{\gamma}G(\by_i;\bth^{\ast})dy_i\\
&+\Big(\prod_{j=1}^{n_i}(1-\zeta_{1ij})\Big)\zeta_{2i}\int\phi(y_{ij};\bx_{ij}^{\top}\bbe+\bz_{ij}^{\top}\bbt_i(\by_i;\bth),\sigma^2)^{\gamma}h_1(\by_i;\bth^{\ast})dy_i\\
&+\zeta_{2i}\int\phi(y_{ij};\bx_{ij}^{\top}\bbe+\bz_{ij}^{\top}\bbt_i(\by_i;\bth),\sigma^2)^{\gamma}\sum_{s_i\in D_i}h_2(\by_i;\bth^{\ast})dy_i\\
&+(1-\zeta_{2i})\int\phi(y_{ij};\bx_{ij}^{\top}\bbe+\bz_{ij}^{\top}\bbt_i(\by_i;\bth),\sigma^2)^{\gamma}\sum_{s_i\in D_i}h_3(\by_i;\bth^{\ast})dy_i,
\end{align*}
noting that $\zeta_i$ is the marginal contamination probability defined in (C3).

From the definition of $\Omega_\nu$, the above formula can be evaluated from the upper by 
\begin{align*}
&(1-\zeta_i)\int\phi(y_{ij};\bx_{ij}^{\top}\bbe+\bz_{ij}^{\top}\bbt_i(\by_i;\bth),\sigma^2)^{\gamma}G(\by_i;\bth^{\ast})dy_i+\nu.
\end{align*}
Then, for $\bth\in \Omega_\nu$, we have 
\begin{align*}
&\log\left(\frac1{m\bn}\sum_{i=1}^m\sum_{j=1}^{n_i}\int\phi(y_{ij};\bx_{ij}^{\top}\bbe+\bz_{ij}^{\top}\bbt_i(\by_i;\bth),\sigma^2)^{\gamma}f_c(\by_i;\bth^{\ast})dy_i\right) \\
& \leq 
\log\left(\frac1{m\bn}\sum_{i=1}^m\sum_{j=1}^{n_i}(1-\zeta_i) \int\phi(y_{ij};\bx_{ij}^{\top}\bbe+\bz_{ij}^{\top}\bbt_i(\by_i;\bth),\sigma^2)^{\gamma}G(\by_i;\bth^{\ast})dy_i+\nu\right)\\
&=\log\left(\frac1{m\bn}\sum_{i=1}^m\sum_{j=1}^{n_i}(1-\bzeta)\int\phi(y_{ij};\bx_{ij}^{\top}\bbe+\bz_{ij}^{\top}\bbt_i(\by_i;\bth),\sigma^2)^{\gamma}G(\by_i;\bth^{\ast})dy_i+\nu+O_p\left(\frac1m\right)\right),
\end{align*}
where $\bzeta=E(\zeta_i)$.
Then, the above can be simplified to
\begin{align}
&\log\left(\frac1{m\bn}\sum_{i=1}^m\sum_{j=1}^{n_i}(1-\bzeta)\int\phi(y_{ij};\bx_{ij}^{\top}\bbe+\bz_{ij}^{\top}\bbt_i(\by_i;\bth),\sigma^2)^{\gamma}G(\by_i;\bth^{\ast})dy_i\right)\notag\\
& \ \ \ \ \ \ \ 
+\left(\nu+O_p\left(\frac1m\right) \right)C(\by;\bth)+O(\nu^2)+O_p\left(\frac1m\right)\notag\\
&=\log\left(\frac1{m\bn}\sum_{i=1}^m\sum_{j=1}^{n_i}\int\phi(y_{ij};\bx_{ij}^{\top}\bbe+\bz_{ij}^{\top}\bbt_i(\by_i;\bth),\sigma^2)^{\gamma}G(\by_i;\bth^{\ast})dy_i\right)\notag\\
& \ \ \ \ \ \ \ 
+\nu C_1(\by;\bth)+O(\nu^2)+O_p\left(\frac1m\right)+\log(1-\bzeta), \label{eq:C1}
\end{align}
where 
\begin{align*}
&C_1(\by;\bth)=\left(\frac1{m\bn}\sum_{i=1}^m\sum_{j=1}^{n_i}\int\phi(y_{ij};\bx_{ij}^{\top}\bbe+\bz_{ij}^{\top}\bbt_i(\by_i;\bth),\sigma^2)^{\gamma}G(\by_i;\bth^{\ast})dy_i\right)^{-1}.
\end{align*}
Note that (\ref{eq:C1}) holds uniformly on $\bth\in \Omega_{\nu}$ since $\Omega_\nu$ is compact and 
\begin{align*}
&\frac1{m\bn}\sum_{i=1}^m\sum_{j=1}^{n_i}\int\phi(y_{ij};\bx_{ij}^{\top}\bbe+\bz_{ij}^{\top}\bbt_i(\by_i;\bth),\sigma^2)^{\gamma}G(\by_i;\bth^{\ast})dy \\
&\leq \frac1{m\bn}\sum_{i=1}^m\sum_{j=1}^{n_i}\int G(\by_i;\bth^{\ast})dy=\frac1{m\bn}\sum_{i=1}^m n_i<\infty.
\end{align*}
Applying the same evaluation, we get 
\begin{align}
&\log\left(\frac1m\sum_{i=1}^m\int\phi_q(\bbt_i(\by_i;\bth);\zero,\bR)^{\gamma}f_c(\by_i;\bth^{\ast})dy_i\right)\notag\\
\leq &\log\left(\frac1m\sum_{i=1}^m\int\phi_q(\bbt_i(\by_i;\bth);\zero,\bR)^{\gamma}G(\by_i;\bth^{\ast})dy_i\right)\notag\\
& \ \ \ + \nu C_2(\by,\bth)+O(\nu^2)+O_p\left(\frac1m\right)+\log(1-\bzeta),\label{eq:C2}
\end{align}
where 
$$
C_2(\by,\bth)=\left(\frac1m\sum_{i=1}^m\int\phi_q(\bbt_i(\by_i;\bth);\zero,\bR)^{\gamma}G(\by_i;\bth^{\ast})dy_i\right)^{-1}.
$$
Combining (\ref{eq:C1}) and we finally have 
\begin{equation}\label{D_gamma}
m^{-1}D_{\gamma}^{M\dagger}(\bth)
\leq m^{-1}D_{\gamma}^{M\ast}(\bth)+\nu C^{\ast}(y;\bth)+O(\nu^2)+O_p\left(\frac1m\right)+2\log(1-\bzeta),
\end{equation}
uniformly on $\bth\in\Omega_\nu$, where $C^{\ast}(\by;\bth)=\gamma^{-1}\bn C_1(\by,\bth) + \gamma^{-1}C_2(\by,\bth)$. 
The relation (\ref{D_gamma}) between $m^{-1}D_{\gamma}^{M\dagger}(\bth)$ and $m^{-1}D_{\gamma}^{M\ast}(\bth)$ shows that $\bth_{\gamma}^{\dagger}=\bth_{\gamma}^{\ast}+O(\nu)$.

\end{proof}

\section{Additional simulation results}

\subsection{Results of error variances in Section~4}
\begin{figure}[!htb]
\centering
\includegraphics[width=10cm,clip]{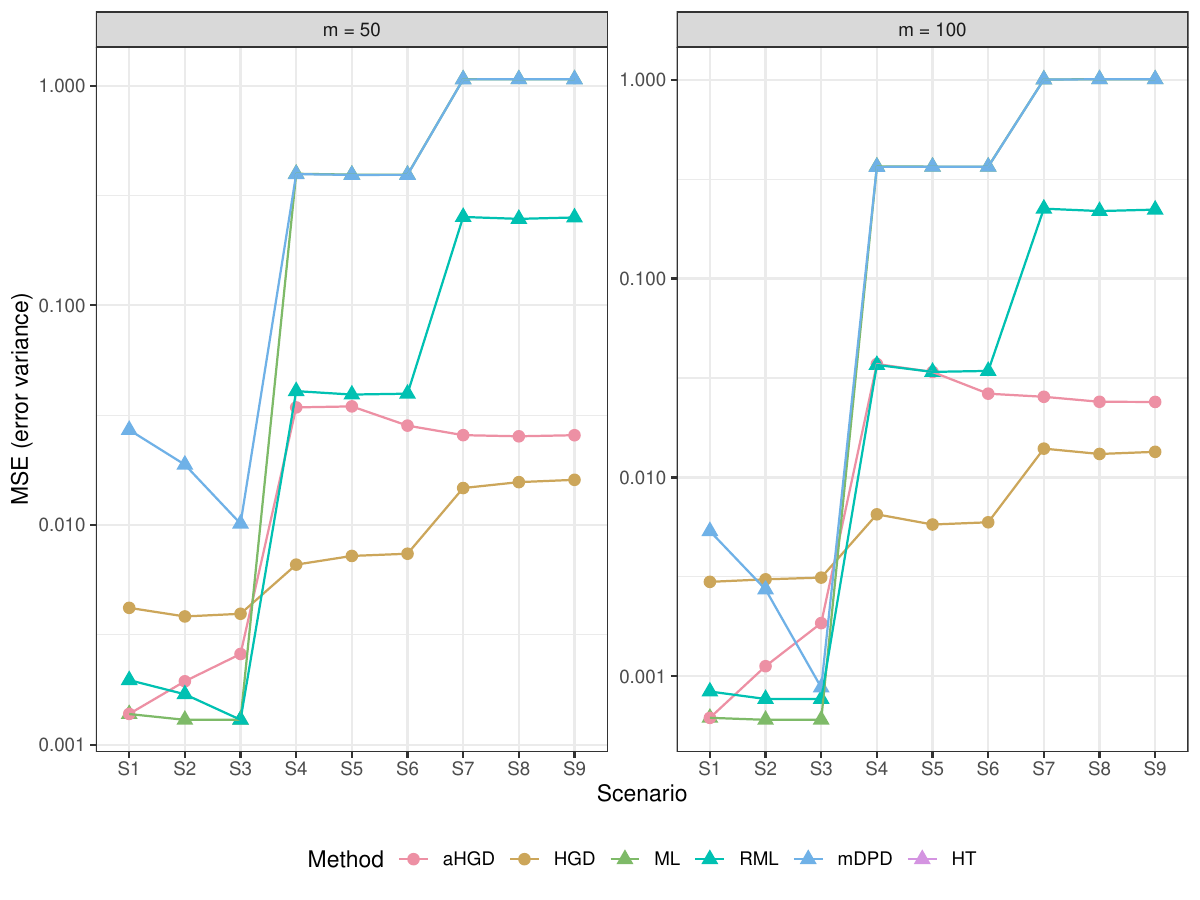}
\caption{
Mean squared errors (MSE) for estimators of error variance $\sigma$ across nine different contamination scenarios under $m=50$ and $m=100$.
\label{fig:sim-supp}
}
\end{figure}

We here present additional simulation results related to Section~4. 
In Figure~\ref{fig:sim-supp}, we present MSE values regarding the error variance under nice scenarios. 
As confirmed in Figure~1 in the main document, the proposed aHGD and HGD provides better estimation accuracy than the other methods.

\subsection{Results under alternative scenarios}
In this section, we consider simulation studies with different data generating processes from those in Section~4. The same settings for the number of clusters, cluster sizes, covariate distributions and true parameters were adopted, but critically alternative settings of the true generating distributions for random effects and error terms were used as follows: to generate the two-dimensional random effect $\bb_i=(b_{0i}, b_{1i})^\top$, we first define the random variable $\bb_i^{\ast}$ as 
$$
\bb_i^{\ast}=-\bde \mu +\bde \frac{|u_i|}{\sqrt{w_i}}+\bv_i, \ \ \ \bv_i\sim N(0, \bR/w_i), \ \ \ \mu=\sqrt{\frac{\nu_b}{\pi}}\frac{\Gamma((\nu_b-1)/2)}{{\Gamma(\nu_b/2)}},
$$
where $u_i\sim N(0,1)$, $w_i\sim {\rm Ga}(\nu_b/2, \nu_b/2)$ and $u_i, w_i$ and $\bv_i$ are mutually independent. 
The above specification of $\bb_i$ is the stochastic representation of the restricted multivariate skew-$t$ distribution \citep[e.g.,][]{lee2016finite}, with the parameters $\nu_b$ and $\bde$ controlling the degree of freedom and skewness, respectively. 
Note $E[\bb_i^{\ast}]=0$ and 
$$
{\rm Var}(\bb_i^{\ast})\equiv \bR_{\ast}=\frac{\nu_b}{\nu_b-2}\bR +\left(\frac{\nu_b}{\nu_b-2}-\mu\right)\bde\bde^\top.
$$
We then define the random effect $\bb_i$ as $\bb_i=\bR^{1/2}\bR_{\ast}^{-1/2}\bb_i^{\ast}$, which guarantees that $E[\bb_i]=0$ and ${\rm Var}(\bb_i)=\bR$. 
Moreover, the error term $\ep_{ij}$ is generated from a mixture distribution with density 
$$
f(\ep_{ij})=0.9 \cdot \phi(\ep_{ij}; 0, \sigma^2) + 0.1\cdot t_{\nu_\ep}(\ep_{ij}),
$$
where $t_{\nu_\ep}(\cdot)$ denotes the density function of a $t$-distribution with $\nu_\ep$ degrees of freedom.

\begin{figure}[!htb]
\centering
\includegraphics[width=14cm,clip]{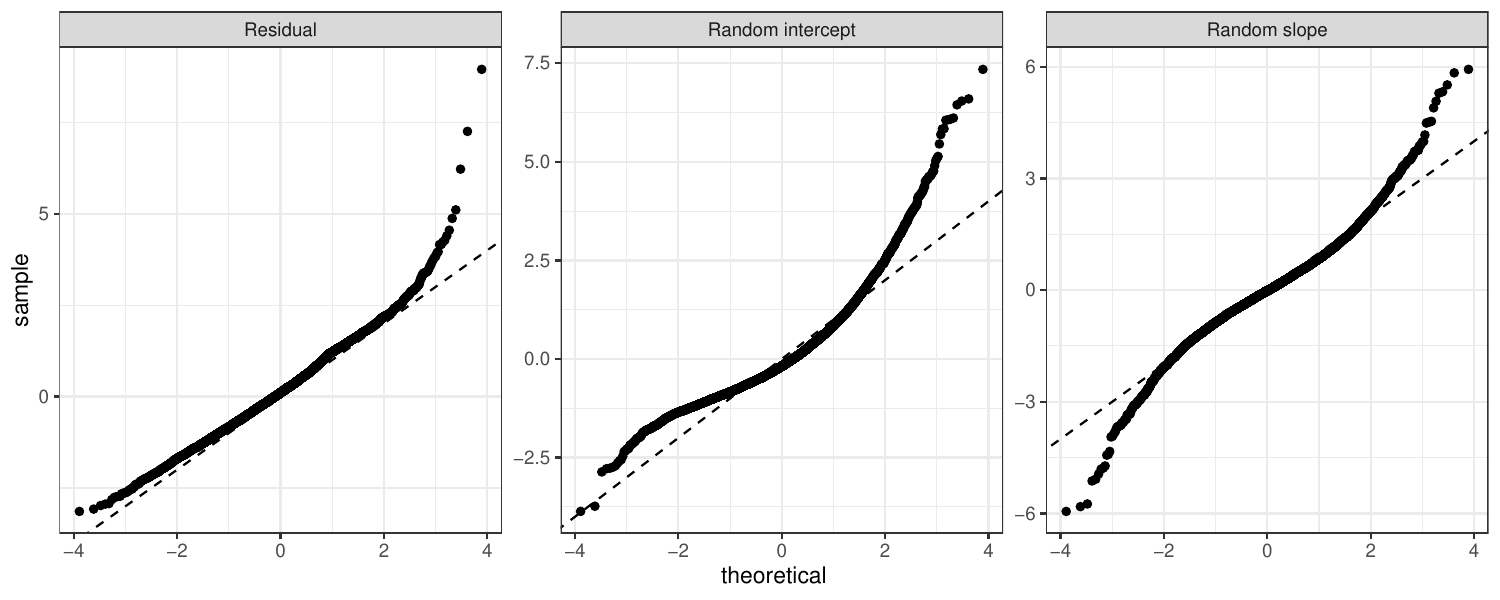}
\caption{Normal probability plots of the generated error term (left), random intercept (center) and random slope (right) under Scenario~(SS1). }
\label{fig:sim-supp-qqplot}
\end{figure}

We adopted the following six scenarios of $\bde$, $\nu_b$ and $\nu_{\ep}$: 
\begin{align*}
&{\rm (SS1)}: \bde=(3,0), \ \nu_b=5,\ \nu_\ep=3, \quad \quad
{\rm (SS2)}: \bde=(0,3), \ \nu_b=5,\ \nu_\ep=3\\
&{\rm (SS3)}: \bde=(3,3), \ \nu_b=5,\ \nu_\ep=3, \quad \quad
{\rm (SS4)}: \bde=(3,0), \ \nu_b=5,\ \nu_\ep=2\\
&{\rm (SS5)}: \bde=(0,3), \ \nu_b=5,\ \nu_\ep=2, \quad \quad
{\rm (SS6)}: \bde=(3,3), \ \nu_b=5,\ \nu_\ep=2.
\end{align*}
The above parameter settings are motivated by the application in Section~5. In particular, we generated 10,000 samples from the true random effects and error distributions, and present the normal probability plots in Figure~\ref{fig:sim-supp-qqplot}.
It can be seen that shape of these distributions are very similar to those observed in Figure~3 in the main text.

\begin{figure}[H]
\centering
\includegraphics[width=10cm,clip]{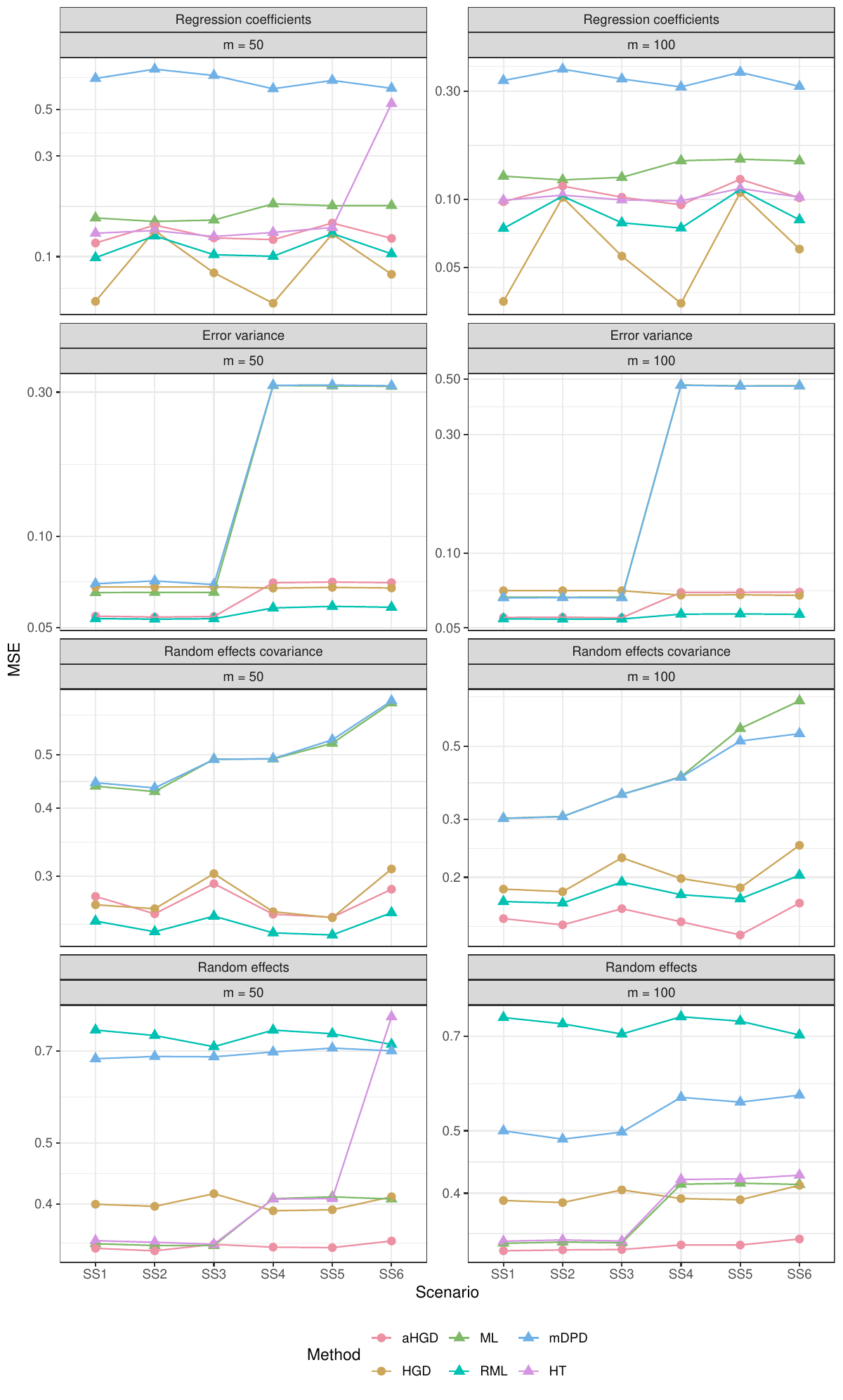}
\caption{Mean squared errors (MSE) for different estimators of the regression coefficients $\bbe$ (the first row), error variance $\sigma^2$ (the second row), random effect covariance matrix $\bR$ (the third row) and random effects $\bb_i$ (the forth row) across six different data generating processes and $m=50$ (left) and $m=100$ (right) clusters. 
\label{fig:sim-mse-supp}
}
\end{figure}

For the generated data, we applied the same methods used in Section 4 of the main text, and evaluated mean squared errors (MSE) of the point estimates of the regression coefficients, error variance, random effects covariance and random effects, based on 500 replicated datasets. The results under six scenarios are presented in Figure~\ref{fig:sim-mse-supp}.
Overall, the proposed HGD approach outperformed ML, mDPD and HT, and was comparable with that of RML in the model parameters, but greatly outperformed RML in terms of random effects prediction.

\section{Application to AIDS Cohort Study with Transformed Response Values}

\begin{figure}[H]
\centering
\includegraphics[width=0.9\textwidth,clip]{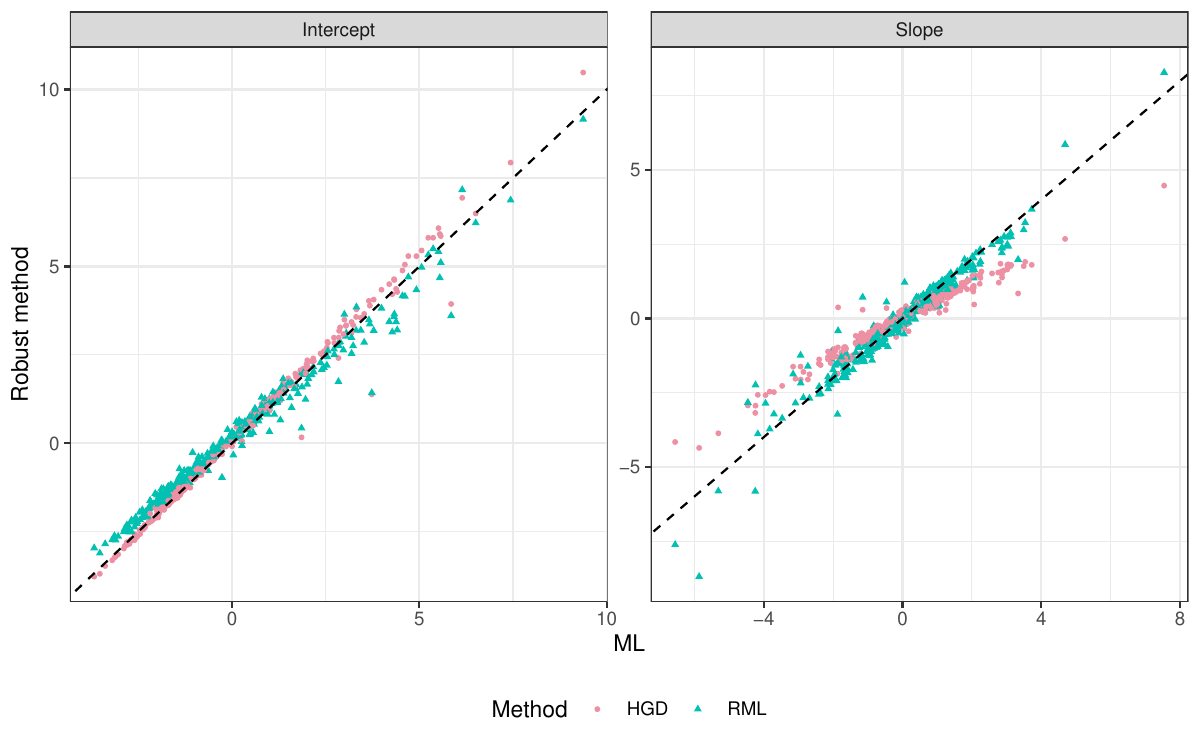}
\caption{Scatter plots of predicted random effects obtained by ML and the two robust methods (RML and HGD).
}
\end{figure}

In Section 5 of the main text, applying linear mixed models to the response variable (the number of CD4 cells) suggested the presence of outliers (as confirmed in Figure~\ref{fig:resid}), and this motivated the use of the proposed HGD method. 
However, the presence of outliers could be due to the misspecification of the shape of error term and/or random effects distribution itself. Therefore, in this section, we apply three different, potentially variance-stabilizing transformations (logarithmic, squared root and cubic root transformations) to the response before fitting the same linear mixed models using ML, HGD, and RML. 

\begin{figure}[!htb]
\centering
\includegraphics[width=14cm,clip]{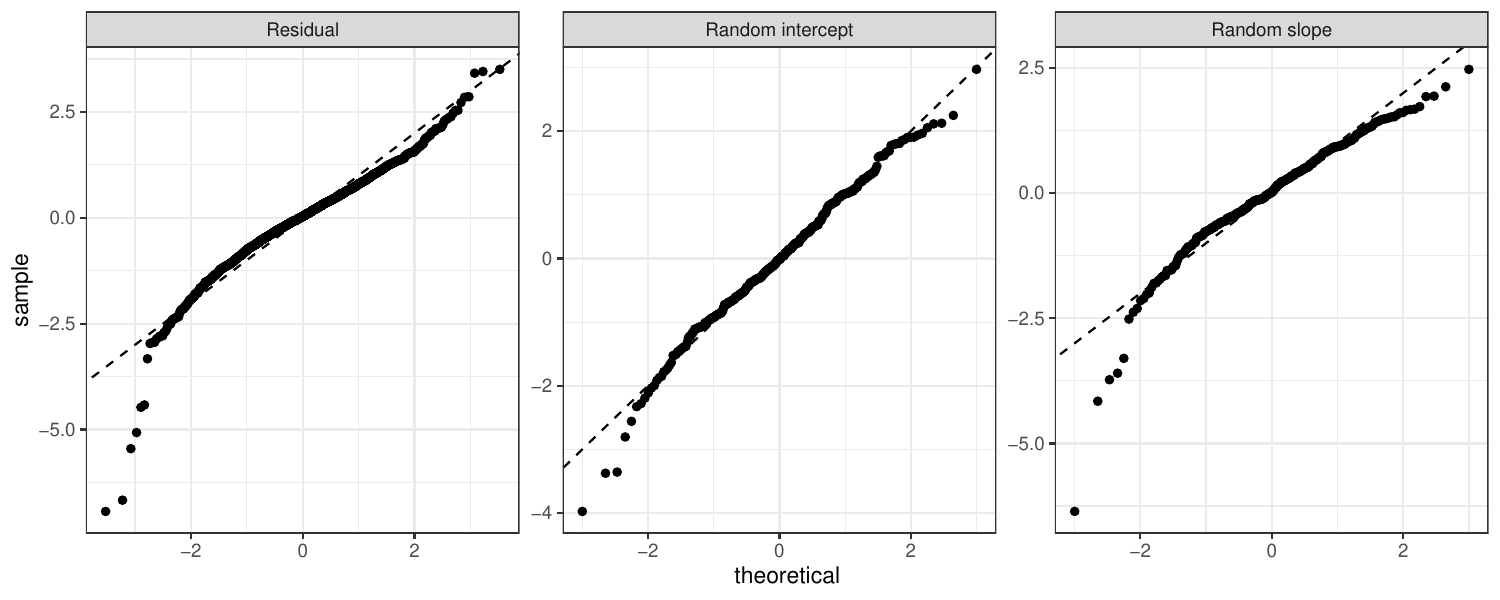}
\includegraphics[width=14cm,clip]{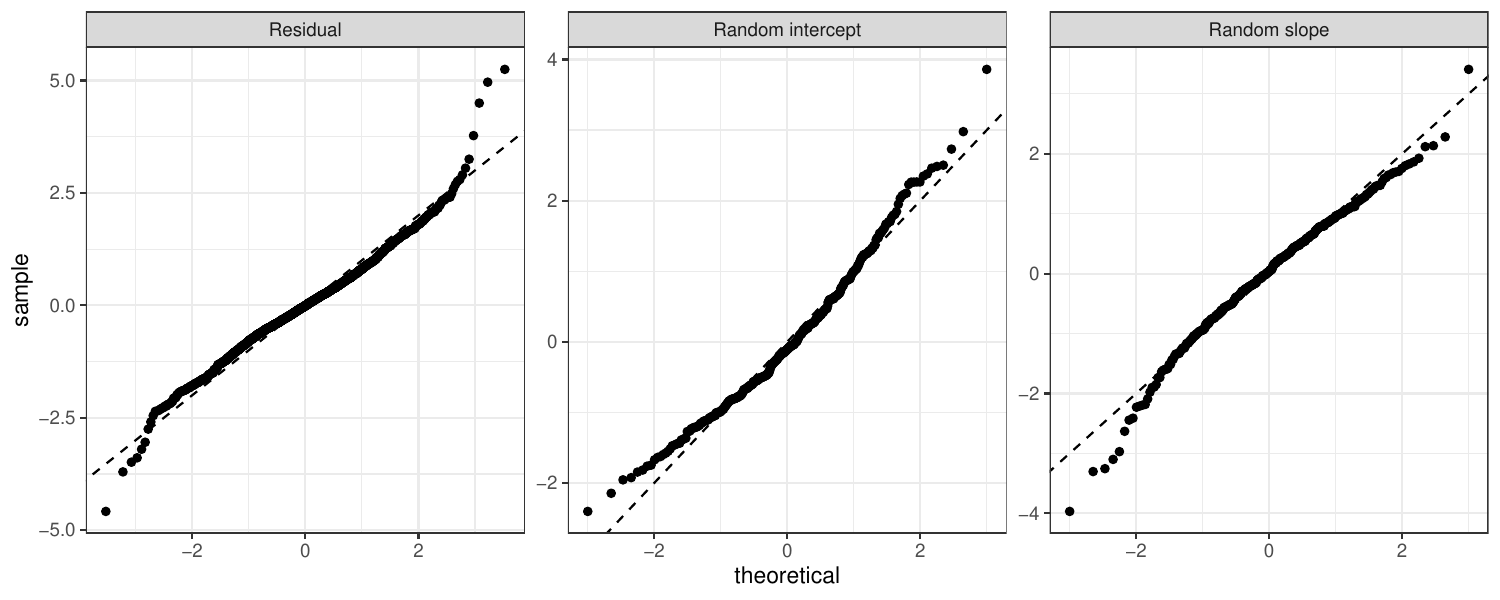}
\includegraphics[width=14cm,clip]{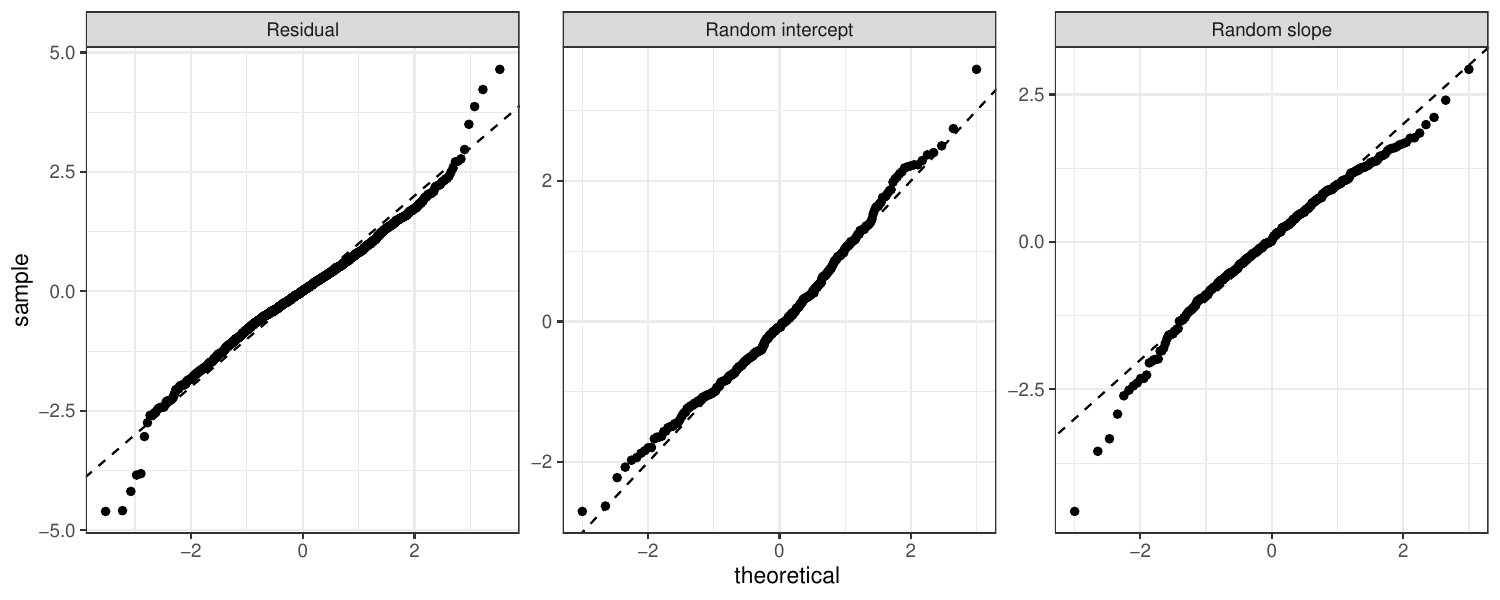}
\caption{ 
Normal probability plots of the standardized residuals (left), random intercept (center)
and random slope (right), based on fitting the LMM to the AIDS cohort study using standard
maximum likelihood estimation under logarithmic (upper), squared root (middle) and cube root (bottom) transformations. 
\label{fig:app-sup1}
}
\end{figure}

\begin{figure}[!htb]
\centering
\includegraphics[width=14cm,clip]{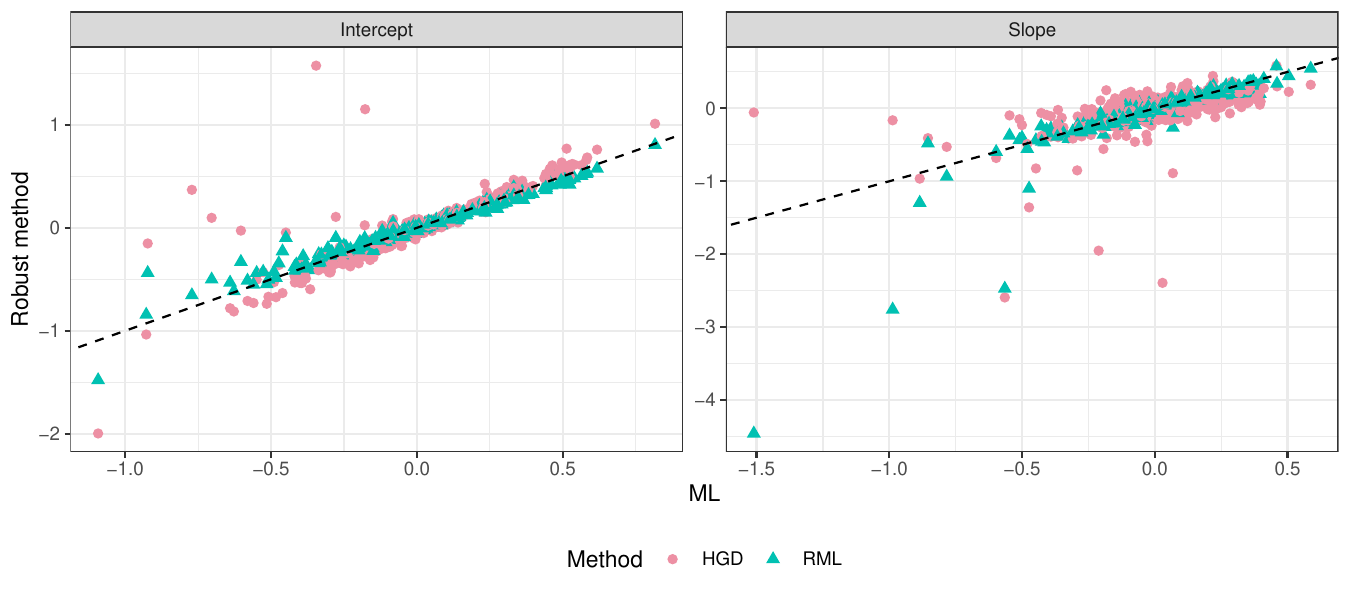}
\includegraphics[width=14cm,clip]{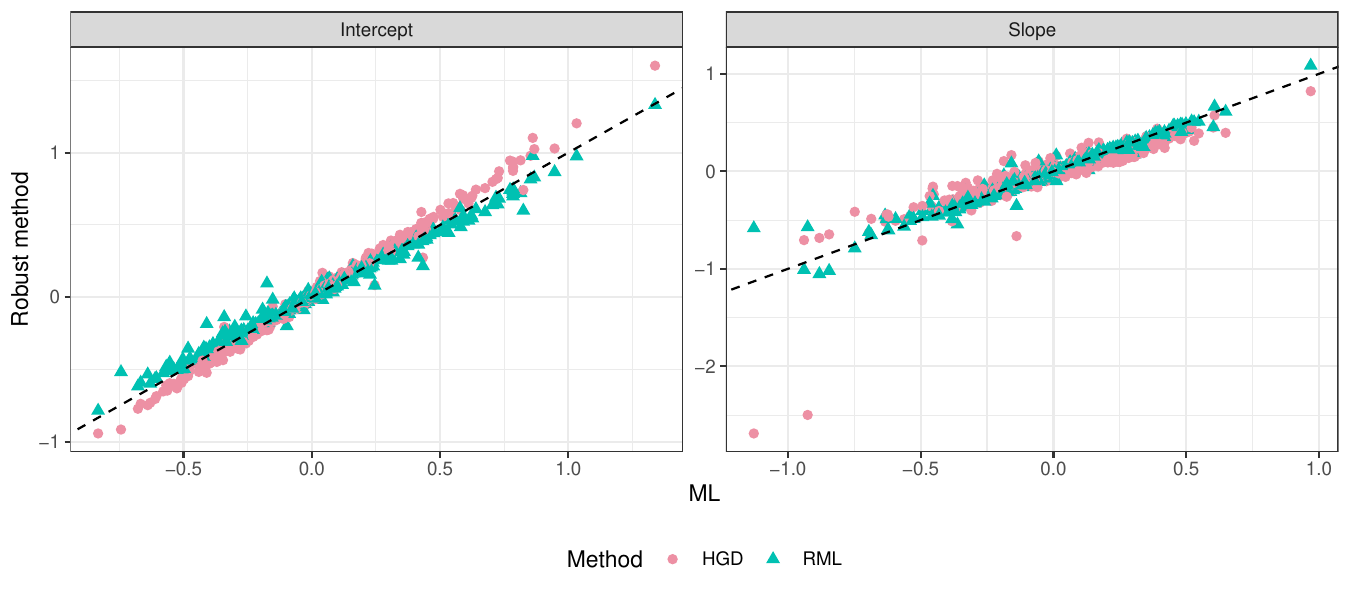}
\includegraphics[width=14cm,clip]{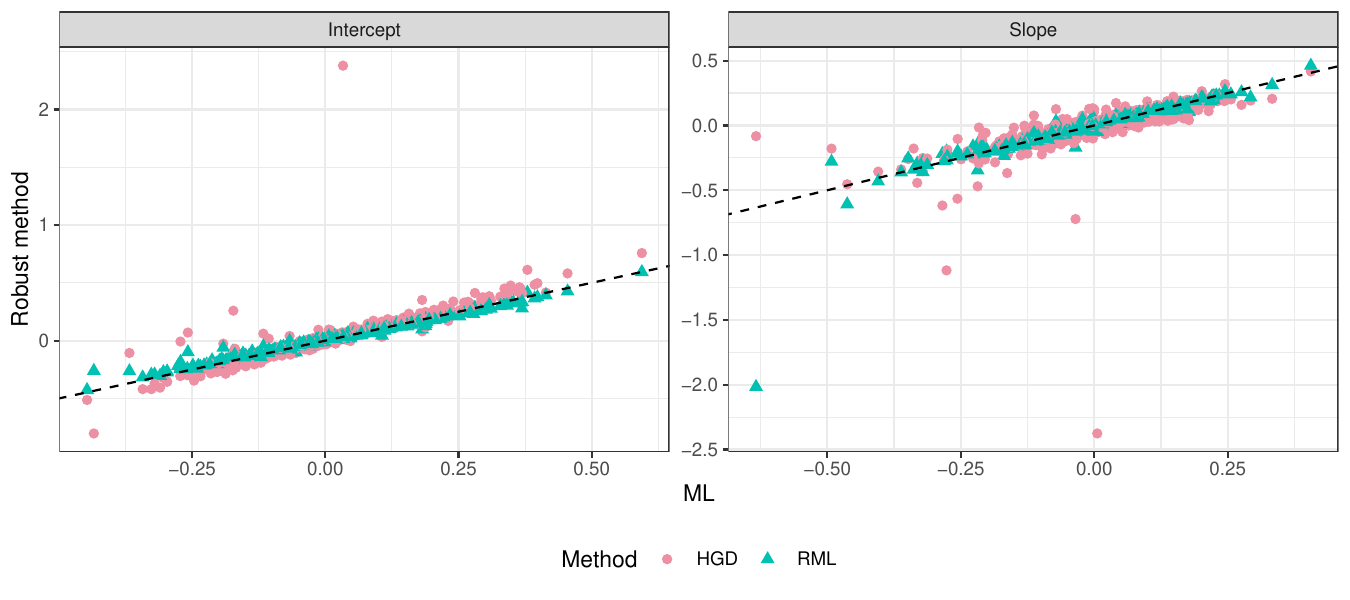}
\caption{
Scatter plots of predicted random effects obtained by ML and the two robust methods (RML and HGD) under logarithmic (upper), squared root (middle) and cube root (bottom) transformations. 
\label{fig:app-sup2}
}
\end{figure}

Figure~\ref{fig:app-sup1} presents normal probability plots of residuals, predicted random intercept, and predicted random slope of the ML method. All indicate that the normality assumption seems to be violated regardless of the transformation used for the response. 
In particular, some outlying observations can be found in the residual plots after all transformations; this suggests that applying transformations is not sufficient to remove the outliers identified in the main analysis given in Section~5.
This is further confirmed when we computed the optimal $\gamma$ for proposed HGD approach via the H-score, and selected among $\{0, 0.01,\ldots, 0.49, 0.50\}$:
$$
\gamma=0.30 \ ({\rm logarithmic}),  \ \ \ \ 
\gamma=0.10 \ ({\rm square\ root}), \ \ \ \ 
\gamma=0.42 \ ({\rm cube\ root}),
$$
All of these are positive, offering some evidence of outliers being present in the data and the need to employ a robust mixed modeling approach. Finally Figure~\ref{fig:app-sup2} provides scatterplots of the predicted random effects obtained by ML and the two robust methods (RML and HGD) under three transformations. All of these suggest that HGD can detect the outlying random effects better than RML under three transformations.

\end{document}